\documentclass{elsart}
\usepackage{graphicx}
\usepackage{amsmath,amsbsy,amssymb,bm}
\numberwithin{equation}{section}

\journal{Annals of Physics}

\newcommand{\intx}[1][]{\ensuremath{\int d^3 x{#1}}}
\newcommand{\intk}{\ensuremath{\int \frac{d^3 k}{(2\pi)^3}}}
\newcommand{\intq}{\ensuremath{\int \frac{d^3 q}{(2\pi)^3}}}
\newcommand{\intomega}{\ensuremath{\int^{+\infty}_{-\infty} d\omega}}
\newcommand{\bx}[1][]{\ensuremath{\mathbf{x}{#1}}}
\newcommand{\bk}{\ensuremath{\mathbf{k}}}
\newcommand{\bp}{\ensuremath{\mathbf{p}}}
\newcommand{\bq}{\ensuremath{\mathbf{q}}}
\newcommand{\xt}[1][]{\ensuremath{\mathbf{x}{#1},t{#1}}}
\newcommand{\kt}{\ensuremath{\mathbf{k},t}}

\newcommand{\komega}{\ensuremath{k,\omega}}
\newcommand{\nn}{\nonumber}

\begin{document}
\begin{frontmatter}
\title{Nonequilibrium relaxation of Bose-Einstein condensates:
Real-time equations of motion and Ward identities}
\author[pitt]{D. Boyanovsky\corauthref{cor}}, \ead{boyan@pitt.edu}
\corauth[cor]{Corresponding author.}
\author[pitt]{S.-Y. Wang\corauthref{ca}}, \ead{swang@lanl.gov}
\corauth[ca]{Current address: Theoretical Division, MS B285, Los
Alamos National Laboratory, Los Alamos, New Mexico 87545, USA.}
\author[donghwa]{D.-S. Lee}, \ead{dslee@mail.ndhu.edu.tw}
\author[sinica]{H.-L. Yu}, \ead{phhlyu@ccvax.sinica.edu.tw}
and
\author[kingfahd]{S. M. Alamoudi} \ead{alamoudi@kfupm.edu.sa}
\address[pitt]{Department of Physics and Astronomy, University of
Pittsburgh, Pittsburgh, Pennsylvania 15260, USA}
\address[donghwa]{Department of Physics, National Dong
Hwa University, Shoufeng, Hualien 974, Taiwan, ROC}
\address[sinica]{Institute of Physics, Academia Sinica, Taipei 115,
Taiwan, ROC}
\address[kingfahd]{Department of Physics, King Fahd
University of Petroleum and Minerals, Dhahran, Saudi Arabia}

\begin{abstract}
We present a field-theoretical method to obtain consistently the
equations of motion for small amplitude condensate perturbations
in a homogeneous Bose-condensed gas directly in real time. It is
based on linear response, and combines the Schwinger-Keldysh
formulation of nonequilibrium quantum field theory with the
Nambu-Gor'kov formalism of quasiparticle excitations in the
condensed phase and the tadpole method in  quantum field theory.
This method leads to causal equations of motion that allow to
study the nonequilibrium evolution as an initial value problem. It
also allows to extract directly the Ward identities, which are a
consequence of the underlying gauge symmetry and which in
equilibrium lead to the Hugenholtz-Pines theorem. An explicit
one-loop calculation of the equations of motion beyond the
Hartree-Fock-Bogoliubov approximation reveals that the nonlocal,
absorptive contributions to the self-energies corresponding to the
Beliaev and Landau damping processes are necessary to fulfill the
Ward identities \emph{in} or \emph{out} of equilibrium. It is
argued that a consistent implementation at low temperatures must
be based on the loop expansion, which is shown to fulfill the Ward
identities order by order in perturbation theory.
\end{abstract}
\end{frontmatter}

\section{Introduction}
The realization of Bose-Einstein condensation (BEC)
\cite{bec1,bec2,bec3,ketterle:rev} in dilute atomic gases in which
the atoms are confined in magnetic traps and cooled down to
temperatures of order of fraction of microkelvins has opened a new
era in atomic physics. The experimental and theoretical effort on
the understanding of BEC is truly interdisciplinary~\cite{snoke}
and the beautiful demonstration of BEC in atomic gases has
rekindled both the experimental and theoretical
interests~\cite{ketterle:rev}.

The theoretical description of Bose-Einstein condensation in
weakly interacting dilute gases has a long
history~\cite{book:griffin,dalfovo99,shi,fetter:art,castin} and
has accounted for many experimental
results~\cite{bec1,bec2,bec3,ketterle:rev}. Pioneering work on the
microscopic theory of a Bose-condensed gas was done by
Bogoliubov~\cite{bogoliubov}, whose original observation that the
wave function of the excitations around the condensate (phonons)
is a linear superposition of free particle states with equal and
opposite momenta has been recently brilliantly confirmed
experimentally~\cite{bogoexp}. A detailed and comprehensive
theoretical description of the homogeneous and inhomogeneous BEC
is offered in
Refs.~\cite{book:griffin,dalfovo99,shi,fetter:art,snoke}.

The current experiments on trapped Bose-condensed gases can
measure with great accuracy the dynamical aspects of the
collective (quasiparticle) excitations (see
Ref.~\cite{ketterle:rev} for a thorough discussion). The earlier
experiments~\cite{bec1,bec2,bec3} were carried at low temperatures
and the measurements of the dynamics of the condensate revealed
almost undamped oscillations with frequencies in excellent
agreement with theoretical predictions based on the time dependent
Gross-Pitaevskii
equation~\cite{GP,book:griffin,dalfovo99,shi,fetter:art,griffin96,giorgini98}.

More recent experiments~\cite{bec3,ketterle:rev} probe the low
energy dynamics of collective excitations at higher temperatures
and show evidence for large frequency shifts and strong damping.
There are several important processes that contribute to damping
of collective excitations. At zero temperature a quasiparticle can
decay into two or more quasiparticles of lower energies, a
mechanism that was originally studied by Beliaev~\cite{beliaev} at
zero temperature and by Popov~\cite{popovld} at finite temperature
in a homogeneous Bose gas. At finite temperature an important
damping mechanism in the collisionless regime is the Landau
damping~\cite{liu,fed,pita}, which has recently been studied for
trapped Bose gases~\cite{giorgini98} within a time dependent
mean-field approximation.

The necessity for a firm theoretical understanding of the
dynamical aspects of low-lying collective excitations in the
condensed phase is underscored by the intense experimental effort
to probe them. For a homogeneous Bose gas in the condensed phase a
classification scheme for the different types of theoretical
approximations to study the excitations has been put forth in a
seminal article by Hohenberg and Martin~\cite{hohenberg} and more
recently discussed in detail by
Griffin~\cite{book:griffin,griffin96,shi}. A very important
ingredient for a consistent description of the dynamics of
quasiparticle excitations is the Hugenholtz-Pines
theorem~\cite{hugenholtz}, which is a consequence of the
underlying gauge invariance~\cite{hohenberg} and similar to the
Goldstone theorem in that it guarantees gapless single
(quasi)particle excitations if a continuous symmetry is broken
with short range forces~\cite{peskin}. Since the Hugenholtz-Pines
theorem is a consequence of the original gauge invariance and the
Ward identities that stem from it, it is important to guarantee
that any approximation scheme to study the dynamics of low energy
excitations respects this theorem~\cite{griffin96}. Detailed
studies of the equations of motion for the small amplitude
perturbation in the condensed phase~\cite{griffin96,giorgini98}
highlight the necessity for a systematic treatment compatible with
the Hugenholtz-Pines theorem. The usual approach to obtaining the
equations of motion begins with taking the expectation value of
the Heisenberg equations of motion and invoking some type of
factorization of the nonlinear correlators (see, e.g.,
Refs.~\cite{griffin96,giorgini98} and references therein). It is
at this point where approximation schemes to treat the nonlinear
correlator may potentially conflict with the Hugenholtz-Pines
theorem and the Ward identities.

\textbf{The focus and goals.} We focus on studying the real-time
evolution of the condensate perturbations in a weakly interacting,
homogeneous Bose-condensed gas to begin with, but our ultimate
goals are to study consistently the nonequilibrium evolution of
collective excitations and to establish contact with the
experimental effort. In particular, we seek to obtain the
oscillation frequencies and damping rates of condensate
perturbation both for the homogeneous as well as for the
inhomogeneous (trapped) cases and to study the nonequilibrium
aspects of quasiparticle excitations in a Bose-condensed gas.

While the study of the homogeneous Bose gas may not be directly
related to the experimental effort, a deeper understanding of
nonequilibrium phenomena in many-body systems is intrinsically of
fundamental importance and may pave the way to a systematic
treatment of similar effects in trapped atomic gases. Of
particular relevance for our study, at least for the homogeneous
Bose gas, is to assess the impact of the infrared divergences in
the self-energies on the frequencies and damping rates of
collective excitations. In a strict perturbative expansion in the
weakly interacting homogeneous Bose gas, the quasiparticle
self-energies feature infrared divergences, logarithmic at zero
temperature and power law at finite
temperature~\cite{book:griffin,shi}. Early work by Gavoret and
Nozi\`{e}res~\cite{gavoret}, Nepomnyashchy and
Nepomnyashchy~\cite{nepomnyashchy} and
Popov~\cite{popov,book:popov} points out that a resummation of the
perturbative expansion leads to a cancellation of the infrared
divergences in physical quantities. This result in turn suggests
that the perturbative expansion requires a resummation to extract
the quasiparticle frequencies and damping rates unambiguously, but
the resummation scheme invoked should fulfill the Hugenholtz-Pines
theorem. To our knowledge these issues have not been so thoroughly
studied and whether such divergences require novel treatment in
the case of trapped atomic gases seems to be an open question.

In this article we propose a consistent formulation of the
nonequilibrium dynamics by implementing a field-theoretical
approach to study dynamics out of equilibrium. The
field-theoretical approach to BEC has been clearly formulated both
in the operator~\cite{book:fetter} as well as in the path integral
formulation~\cite{book:popov,stoofinst}. The functional integral
formulation is particularly convenient as it can be generalized to
study nonequilibrium phenomena by implementing the
Schwinger-Keldysh
formulation~\cite{schwinger,keldysh,kt,chou,mahan,kine} at the
level of the path integrals. Recently, Stoof~\cite{stoof:ft} has
advanced the Schwinger-Keldysh formulation of nonequilibrium field
theory to study the dynamics of condensation in Bose-condensed
gases.

Our ultimate goal is to obtain a consistent and systematic
analysis of quasiparticle frequencies and damping rates in both
the homogeneous and inhomogeneous Bose-condensed gases from a
nonequilibrium approach. We begin such program in this article by
establishing a nonequilibrium framework that allows to extract
this information directly from the real-time evolution of small
amplitude perturbations around the condensate in a homogeneous
Bose-condensed gas.

\textbf{Brief summary of results.} The main results of this
article, which we consider the first in a program that is devoted
to setting up the nonequilibrium formulation, are the summarized
as follows.
\begin{itemize}
\item[(i)]{We combine the Schwinger-Keldysh nonequilibrium
formulation~\cite{schwinger,keldysh,kt,chou,mahan,kine} and the
Nambu-Gor'kov formalism~\cite{book:griffin,shi,book:fetter,nambu}
for treating the Bogoliubov quasiparticles in the condensed phase,
along with a novel method~\cite{tadpole} introduced within the
context of quantum field theory, to obtain the equations of motion
for small amplitude perturbations away from equilibrium around the
condensate \emph{directly in real time}. This method leads to
causal equations of motion and allows to study the evolution of
small amplitude perturbations of the condensate  as an
\emph{initial value problem}~\cite{ivp} which describes an
experimental situation.}
\item[(ii)]{The novel method~\cite{tadpole} implemented to obtain
directly the equations of motion in real time leads to a simple
derivation of the Ward identities associated with the underlying
gauge symmetry \emph{in} or \emph{out} of equilibrium. In
equilibrium these Ward identities lead directly to the
Hugenholtz-Pines theorem. This method is therefore seen to lead to
consistent equations of motion for quasiparticle excitations of
the condensate, that fulfill the Hugenholtz-Pines theorem when the
homogeneous condensate is in equilibrium. We show explicitly to
one-loop order that the inclusion of absorptive contributions
beyond the Hartree-Fock-Bogoliubov approximation are necessary to
fulfill the Ward identities, which, when the condensate is in
equilibrium become the Hugenholtz-Pines theorem. We obtain
explicitly the one-loop self-energies including the absorptive
contributions and their spectral representations beyond the
Hartree-Fock-Bogoliubov approximation.}
\item[(iii)]{We discuss nonequilibrium aspects such as the instabilities
in the quasiparticle spectrum when the homogeneous condensate is
not in equilibrium. We also discuss subtleties associated with the
perturbative expansion that require a consistent rearrangement of
the perturbative series in a \emph{loop} expansion, not in an
expansion in the bare coupling constant.}
\end{itemize}

The article is organized as follows. In Sec.~\ref{sec:II} we
introduce the model and the linear response formulation to obtain
the equations of motion. In Sec.~\ref{sec:noneq} we introduce the
Schwinger-Keldysh formulation in the Nambu-Gor'kov formalism to
study the nonequilibrium aspects of Bogoliubov quasiparticles. In
section \ref{sec:eqnofmot} we introduce the tadpole method, obtain
the equations of motion directly in real time and cast them in
terms of an initial value problem. We obtain explicitly the
retarded self-energies up to one-loop order and obtain their
spectral representations. In Sec.~\ref{sec:HP} we derive the
generalized Ward identities, present an alternative derivation of
the Hugenholtz-Pines theorem and confirm that the one-loop
equations of motion obtained do fulfill this theorem and highlight
the necessity for including the absorptive parts of the
self-energies. In Sec.~\ref{sec:beyond} we highlight relevant
aspects of the perturbative expansions and describe the condensate
instabilities featured in the quasiparticle spectrum when the
condensate is away from equilibrium. Section~\ref{sec:conclusions}
presents our conclusions and poses new questions. An appendix is
devoted to an alternative derivation of the Bogoliubov
transformation, which facilitates the Schwinger-Keldysh
nonequilibrium formulation.

\section{Preliminaries}\label{sec:II}
\subsection{The model}
The Hamiltonian of a Bose gas in the absence of a trapping
one-body potential is given by
\begin{eqnarray}
H&=&\intx\,\psi^\dagger(\xt)
\left(-\frac{\nabla^2}{2m}\right)\psi(\xt)\nn\\
&&+\,\frac{g}{2}\intx\,\psi^\dagger(\xt)
\psi^\dagger(\xt)\psi(\xt)\psi(\xt),\label{H}
\end{eqnarray}
where $\psi(\xt)$ is the Heisenberg complex scalar field
representing spinless bosons of mass $m$, $g=4\pi a/m$ is the
strength of the pseudopotential with $a>0$ being the $s$-wave
scattering length and we have set $\hbar=k_B=1$. We will recover
$\hbar$ in Sec.~\ref{sec:beyond}, where we analyze the nature of
the loop expansion. The field $\psi(\xt)$ and its Hermitian
conjugate satisfy equal-time commutation relations
\begin{gather}
[\psi(\xt),\psi^\dagger(\bx',t)]=\delta^{(3)}(\bx-\bx'),\nn\\
[\psi(\xt),\psi(\bx',t)]=[\psi^\dagger(\xt),\psi^\dagger(\bx',t)]=0.
\end{gather}
A more systematic treatment of interactions
requires a T-matrix formulation~\cite{stoof:ft} but in order to
simplify the discussion we will only consider the local $s$-wave
interaction at this stage. It will become clear that the method
can be straightforwardly generalized to include the T-matrix
resummation of the interaction.

The Hamiltonian $H$ is invariant under the $U(1)$ gauge
transformation
\begin{gather}
\psi(\xt)\to e^{i\theta}\psi(\xt),\nn\\
\psi^\dagger(\xt)\to e^{-i\theta}\psi^\dagger(\xt),
\end{gather}
where $\theta$ is a constant phase. A consequence of this $U(1)$
gauge symmetry is conservation of the number of particles. Indeed,
the number operator of particles
\begin{equation}
N=\intx\,\psi^\dagger(\xt)\psi(\xt)
\end{equation}
commutes with $H$ and hence is a constant of motion. However, it
is convenient to work in the grand-canonical ensemble in which the
grand-canonical Hamiltonian is given by
\begin{eqnarray}
K &\equiv& H-\mu N\nn\\
&=&\intx\,\psi^\dagger(\xt)\left(-\frac{\nabla^2}{2m}-
\mu\right)\psi(\xt)\nn\\
&&+\,\frac{g}{2}\intx\,\psi^\dagger(\xt)
\psi^\dagger(\xt)\psi(\xt)\psi(\xt),\label{K}
\end{eqnarray}
where the chemical potential $\mu$ is the Lagrange multiplier
associated with conservation of number of particles. The
corresponding Lagrangian (density) is given by
\begin{equation}
\mathcal{L}[\psi^\dagger,\psi]=
\psi^\dagger\left(\frac{i}{2}\overleftrightarrow{\frac{\partial}{\partial
t}}+\frac{\nabla^2}{2m}+\mu\right)\psi - \frac{g}{2}\psi^\dagger
\psi^\dagger\psi\psi,\label{L}
\end{equation}
where $\overleftrightarrow{\frac{\partial}{\partial
t}}=\overrightarrow{\frac{\partial}{\partial
t}}-\overleftarrow{\frac{\partial}{\partial t}}$.

In a Bose-condensed gas, the condensate plays a crucial
role~\cite{bogoliubov} and hence it is convenient to decompose the
field into the condensate and noncondensate parts~\cite{griffin96}
\begin{equation}
\psi(\xt)=\phi(\xt)+\chi(\xt),\quad\langle \chi(\xt) \rangle
=0,\label{decompose}
\end{equation}
where $\phi(\xt)\equiv\langle\psi(\xt)\rangle$ is referred to in
the literature as the condensate wave function and $\chi(\xt)$ is
the noncondensate operator. In the above expression, $\langle
\mathcal{O}(\xt)\rangle=\mathrm{Tr}[\rho\mathcal{O}(\xt)]/\mathrm{Tr}\rho$
denotes the \emph{expectation value} of the Heisenberg operator
$\mathcal{O}(\xt)$ in the \emph{initial density matrix} $\rho$.
The presence of the condensate $\phi(\xt)\ne 0$ leads to
spontaneous breaking of the $U(1)$ gauge symmetry and results in a
profound consequence for the spectrum of the quasiparticle
excitations.

In the absence of trapping potentials and explicit symmetry
breaking external sources, the condensate is homogeneous (i.e.,
space-time independent) and denoted by $\phi(\xt)=\phi_0$, which
is the situation under consideration in this article.

\subsection{Real-time relaxation in linear response}
The goal in this article is to obtain \emph{directly in real time}
the equations of motion for small amplitude perturbations of the
homogeneous condensate in an initial value problem formulation.
Our strategy to study the relaxation of the condensate
perturbation as an initial value problem begins with preparing a
Bose-condensed gas slightly perturbed away from equilibrium by
applying external source coupled to the field. Once the external
source is switched off, the perturbed condensate must relax
towards equilibrium. It is precisely this \emph{real-time
evolution} of the nonequilibrium condensate relaxation that we aim
to study in this article.

Let $\eta(\xt)$ be an external $c$-number source coupled to the
quantum field $\psi(\xt)$, then the Lagrangian given by (\ref{L})
becomes
\begin{equation}
\mathcal{L}[\psi^\dagger,\psi]\to\mathcal{L}[\psi^\dagger,\psi]
+\psi^\dagger\eta+\eta^\ast\psi.\label{linearresponse}
\end{equation}
The presence of external source will induce a (linear) response of
the system in the form of an induced expectation value
\begin{equation}
\langle \psi(\xt) \rangle_{\eta}= \phi_0+
\delta(\xt),\label{response}
\end{equation}
where $\langle\mathcal{O}(\xt)\rangle_{\eta}$ denotes the
expectation value of $\mathcal{O}(\xt)$ in the presence of
external source and $\delta(\xt)$ is the space-time dependent
perturbation of the homogeneous condensate $\phi_0$ \emph{induced
by the external source}. The linear response perturbation
$\delta(\xt)$ vanishes when the external source $\eta(\xt)$
vanishes at all times.  This is tantamount to decomposing the
field into the homogeneous condensate ($\phi_0$), a small
amplitude perturbation induced by the external source
[$\delta(\xt)$], and the noncondensate part [$\chi(\xt)$] as
\begin{equation}
\psi(\xt)= \phi_0+\delta(\xt)+\chi(\xt),\quad\langle \chi(\xt)
\rangle_{\eta} =0. \label{finshift}
\end{equation}

In linear response theory $\delta(\xt)$ can be expressed in terms
of the \emph{exact} retarded Green's function of the field in the
absence of external source~\cite{book:fetter,ivp}. An
experimentally relevant initial value problem formulation for the
real-time relaxation of the condensate perturbation can be
obtained by considering that the external source is adiabatically
switched on at $t= -\infty$ and switched off at $t=0$, i.e.,
\begin{equation}
\eta(\xt)=\eta(\bx)\,e^{\epsilon t}\,\Theta(-t), \quad\epsilon\to
0^+. \label{extsource}
\end{equation}
The adiabatic switching-on of the external source induces a
space-time dependent condensate perturbation $\delta(\xt)$, which
is prepared adiabatically by the external source with a given
value $\delta (\bx,0)$ at $t=0$ determined by $\eta(\bx)$. For
$t>0$ after the external source has been switched off, the
perturbed condensate will evolve in the absence of any external
source relaxing towards equilibrium. Thus, the external source
$\eta(\xt)$ is necessary for preparing an initial state at $t=0$
and setting up an initial value problem. This method has been
applied to study a variety of damping and relaxation phenomena in
relativistic hot and dense plasmas~\cite{ivp} and damping of
photons in a strong magnetic field~\cite{mikheev}.

Using the decomposition (\ref{finshift}) and consistent with
linear response by keeping only the linear terms in $\delta$ and
$\delta^\ast$, the Lagrangian $\mathcal{L}$ becomes (in the
presence of external source)
\begin{equation}
\mathcal{L}[\chi^\dagger,\chi]=\mathcal{L}_0[\chi^\dagger,\chi]
+\mathcal{L}_\mathrm{int}[\chi^\dagger,\chi],\label{L1}
\end{equation}
with
\begin{eqnarray}
\mathcal{L}_0[\chi^\dagger,\chi]&=&
\chi^\dagger\bigg(\frac{i}{2}\overleftrightarrow{\frac{\partial}{\partial
t}}+\frac{\nabla^2}{2m}+\mu-2
g|\phi_0|^2\bigg)\chi-\frac{g}{2}\left(\phi^2_0\chi^\dagger\chi^\dagger
+\phi^{\ast 2}_0\chi\chi\right),\nn\\
\mathcal{L}_\mathrm{int}[\chi^\dagger,\chi]&=&\chi^\dagger
\bigg[\left(i\frac{\partial}{\partial t}+\frac{\nabla^2}{2m}+\mu-2
g |\phi_0|^2\right)\delta-g \phi^{2}_0\delta^\ast\nn\\
&&+\,\phi_0\left(\mu-g|\phi_0|^2\right)+\eta\bigg]-2 g \phi_0^*
\delta \chi^\dagger\chi -g \phi_0\delta
\chi^\dagger\chi^\dagger\nn\\
&&-\,g\phi_0\chi \chi^\dagger\chi^\dagger-g\delta\chi
\chi^\dagger\chi^\dagger -\frac{g}{2} \chi^{\dagger
}\chi^{\dagger}\chi\chi+\mathrm{H.c.}, \label{Lint}
\end{eqnarray}
where we have discarded the $c$-number (field operators
independent) and surface terms. Consistent with linear response,
we have only kept linear terms in $\delta$, $\delta^\ast$, which
are the small amplitude departure from the homogeneous condensate
induced by the external source $\eta$. We note that the Lagrangian
$\mathcal{L}[\chi^\dagger,\chi]$ is obviously invariant under the
gauge transformations
\begin{gather}
\phi_0, \delta,  \chi, \eta \to e^{i\theta} \phi_0 ,
e^{i\theta} \delta, e^{i\theta}\chi,  e^{i\theta}\eta, \nn\\
\phi^{\ast}_0 , \delta^{\ast},  \chi^{\dagger}, \eta^\dagger \to
e^{-i\theta} \phi^{\ast}_0, e^{-i\theta} \delta^{\ast},
e^{-i\theta}\chi^\dagger, e^{-i\theta}\eta^\dagger, \label{gauge}
\end{gather}
which, as will be seen below, is at the heart of the Ward
identities that will lead to the Hugenholtz-Pines theorem.

Whereas in general a gauge transformation is invoked (correctly)
to fix the condensate $\phi_0$ to be \emph{real} for convenience,
this choice corresponds to \emph{fixing a particular gauge}, which
in turn hides the underlying gauge symmetry. In order to obtain
the Ward identity associated with this symmetry we will keep a
complex condensate $\phi_0$ and analyze in detail the
transformation laws of the various contributions to the equations
of motion.

\section{Nonequilibrium formulation}\label{sec:noneq}
\subsection{Generating functional}
The general framework to study of nonequilibrium phenomena is the
Schwinger-Keldysh
formulation~\cite{schwinger,keldysh,kt,chou,mahan,kine}, which we
briefly review here in a manner leads immediately to a path
integral formulation. For an alternative presentation see, e.g.,
Refs.~\cite{stoof:ft}.

Consider that the system is described by an initial density matrix
$\rho$ and a perturbation is switched on at a time $t_0$, so that
the total Hamiltonian for $t>t_0$, $H(t)$, does not commute with
the initial density matrix. The expectation value of a Heisenberg
operator $\mathcal{O}(t)= U^{-1}(t,t_0)\mathcal{O}(t_0) U(t,t_0)$
is given by
\begin{equation}
\langle \mathcal{O}(t)\rangle = \frac{\mathrm{Tr}\rho
U^{-1}(t,t_0)\mathcal{O} U(t,t_0)}{\mathrm{Tr}\rho},
\label{expval}
\end{equation}
where $U(t,t_0)$ is the unitary time evolution operator in the
Heisenberg picture
\begin{equation}
U(t,t_0) = T \exp\left[-i \int_{t_0}^t dt'\,H(t')\right],
\label{evolop}
\end{equation}
with $T$ the time-ordering symbol. If the initial density matrix
$\rho$ describes a state in thermal equilibrium at inverse
temperature $\beta$ with the unperturbed Hamiltonian $H(t<t_0)=H$,
i.e.,
\begin{equation}
\rho = e^{-\beta H} = U(t_0-i\beta,t_0),
\end{equation}
then the expectation value (\ref{expval}) can be written in the
form
\begin{equation}
\langle \mathcal{O}(t)\rangle =
\frac{\mathrm{Tr}U(t_0-i\beta,t_0)U^{-1}(t_0,t)\mathcal{O}
U(t,t_0)}{\mathrm{Tr}U(t_0-i\beta,t_0)}. \label{expval2}
\end{equation}
The numerator of this expression has the following interpretation:
evolve in time from $t_0$ up to $t$, insert the operator
$\mathcal{O}$, evolve back from $t$ to the initial time $t_0$ and
down the imaginary axis in time from $t_0$ to $t_0-i\beta$. The
denominator describes the evolution in imaginary time which is the
familiar description of a thermal density matrix. We note that
unlike the $S$-matrix elements or transition amplitudes,
expectation values of Heisenberg operators require evolution
\emph{forward} and \emph{backward} in time (corresponding to the
$U$ and $U^{-1}$ on each side of the operator $\mathcal{O}$).

The time evolution operators have a path-integral
representation~\cite{negele} in terms of the Lagrangian, and the
insertion of operators can be systematically handled by
introducing sources coupled linearly to the field
operators~\cite{book:kadanoff}, i.e.,
\begin{equation}
\mathcal{L}[\chi^\dagger,\chi]\to
\mathcal{L}[\chi^\dagger,\chi]+j^\ast \chi+\chi^\dagger j.
\label{sources}
\end{equation}
The introduction of sources also allows a systematic perturbative
expansion, since in such an expansion, powers of operators are
obtained by functional derivatives with respect to these sources,
which are set to zero after functional differentiation. We note
that these sources $j$, $j^\ast$ introduced to generate the
perturbative expansion in terms of functional derivatives with
respect to these, are \emph{different} from the external sources
$\eta$, $\eta^\ast$ introduced in (\ref{linearresponse}) to
generate an initial value problem in linear response and to
displace the condensate from equilibrium.

\begin{figure}[t]
\begin{center}
\includegraphics[width=3.25in,keepaspectratio=true]{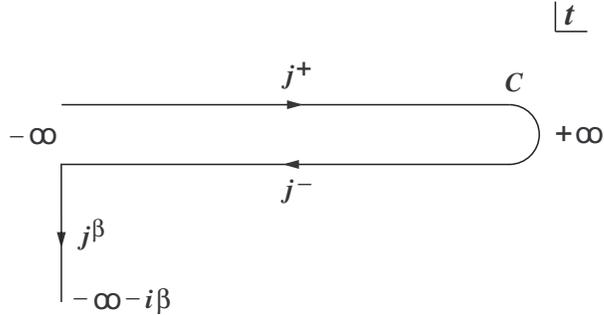}
\caption{The contour $\mathcal{C}$ in complex time plane in the
Schwinger-Keldysh formulation. It consists of a forward branch
running from $t=-\infty$ to $t=+\infty$, a backward branch from
$t=+\infty$ back to $t=-\infty$, and an imaginary branch from
$t=-\infty$ to $t=-\infty-i\beta$. The sources $j^{\pm}$ serve to
generate the real-time nonequilibrium Green's
functions.}\label{fig:ctpcontour}
\end{center}
\end{figure}

Since there are \emph{three} different time evolution operators,
the forward, backward and imaginary, we introduce \emph{three}
different sources for each one of these time evolution operators,
respectively. Taking $t_0 \to -\infty$, we are led to considering
the generating functional~\cite{tadpole,ivp,noneqlect}
\begin{equation}
Z[j^+,j^-,j^\beta]=\mathrm{Tr}U(-\infty-i\beta,-\infty,j^\beta)
U(-\infty,+\infty,j^-)U(+\infty,-\infty;j^+),\label{genefunc}
\end{equation}
where $U(t_f,t_i;j)$ is the time evolution operator [see
\eqref{evolop}] in the presence of the source $j$ and for
simplicity of notation we have not displayed the complex conjugate
of the sources $j^\ast$. The denominator in (\ref{expval2}) is
given by $\mathrm{Tr}\rho = Z[0,0,0]$. The generating functional
$Z[j^+,j^-,j^\beta]$ can be written as a path integral along the
contour in (complex) time plane (see Fig.~\ref{fig:ctpcontour})
\begin{equation}
Z[j^+,j^-,j^\beta] = \int
\mathcal{D}_\mathcal{C}\chi^\dagger\mathcal{D}_\mathcal{C}
\chi\,\exp\left[i\int_\mathcal{C}d^4x\,
\mathcal{L}_\mathcal{C}[\chi^\dagger,\chi,j]\right],\label{pathint}
\end{equation}
where
$\mathcal{D}_\mathcal{C}\chi^\dagger\mathcal{D}_\mathcal{C}\chi$
denotes the functional integration measure along the contour
$\mathcal{C}$ and
\begin{eqnarray}
\int_\mathcal{C}
d^4x\,\mathcal{L}_\mathcal{C}[\chi^\dagger,\chi,j] &\equiv&
\int_{-\infty}^{+\infty}d^4x\,\mathcal{L}[\chi^{\dagger +},
\chi^+,j^+]-\int_{-\infty}^{+\infty}d^4x\,\mathcal{L}[\chi^{\dagger
-},\chi^-,j^-]\nn\\
&&+\,\int_{-\infty}^{-\infty-i\beta}d^4x\,
\mathcal{L}[\chi^{\dagger\beta},\chi^\beta,j^\beta].
\end{eqnarray}
with
$\int_{-\infty}^{+\infty}d^4x\equiv\intx\int_{-\infty}^{+\infty}dt$,
etc. Because of the trace and the bosonic nature of the operators,
the path integral along the contour $\mathcal{C}$ requires
\emph{periodic boundary conditions} on the fields. The
superscripts $+$ and $-$ refer to fields defined in the upper and
lower branches, respectively, corresponding to forward ($+$) and
backward ($-$) time evolution, while the superscript $\beta$
refers to the field defined in the vertical branch running down
parallel to the imaginary axis. The negative sign in front of the
action along the backward branch is a result of the fact that
backward time evolution is determined by $U^{-1}(+\infty,-\infty)$
with $U$ the time evolution operator. The contour source $j$ that
enters in the contour Lagrangian $\mathcal{L}_\mathcal{C}$ in
(\ref{pathint}) takes the values of the sources $j^\pm$ and
$j^\beta$ in the respective branches as displayed in
Fig.~\ref{fig:ctpcontour}.

Functional derivatives with respect to the sources in the forward
branch give time-ordered products of operators, those with respect
to the sources in the backward branch give the anti-time-ordered
products of operators, and those with respect to the sources in
the imaginary branch give the usual imaginary-time (Matsubara)
correlation functions. While the sources $j^+$, $j^-$ and
$j^\beta$ introduced to obtain the correlation functions via
functional differentiation are \emph{different} in the different
branches, as they generate the time-ordered, anti-time-ordered and
Matsubara correlation functions, respectively; the external source
$\eta$, the homogeneous condensate $\phi_0$, and the departure
from equilibrium $\delta$ are $c$-numbers and hence treated the
\emph{same} in all branches.

Writing the Lagrangian as a free and an interaction part as
$\mathcal{L}=\mathcal{L}_0+\mathcal{L}_\mathrm{int}$, the
generating functional can be written as a power series expansion
in the interaction part, which in turn can be generated by taking
functional derivatives with respect to the sources $j$, $j^\ast$
by identifying
\begin{gather}
\chi^\pm \rightarrow \mp i\frac{\delta}{\delta j^{\ast \pm}},\quad
\chi^{\dagger\pm} \rightarrow \mp i\frac{\delta}{\delta j^{\pm}},\nn\\
\chi^\beta \to -i\frac{\delta}{\delta j^{\ast\beta}},\quad
\chi^{\dagger\beta} \to -i\frac{\delta}{\delta
j^{\beta}}.\label{funcders}
\end{gather}
As a result, the full generating functional along the contour
$\mathcal{C}$ can be written as
\begin{equation}\label{PTgenfun}
Z[j] = \exp\left\{i\int_\mathcal{C}
d^4x\,\mathcal{L}_{\mathrm{int},\mathcal{C}}\left[-i\frac{\delta}{\delta
j^\ast},-i\frac{\delta}{\delta j}\right]\right\}Z_0[j],
\end{equation}
where free field generating functional $Z_0[j]$ is given by
(\ref{pathint}) with the non-interacting Lagrangian
$\mathcal{L}_0[\chi^\dagger,\chi]$ given by (\ref{Lint}).

\subsection{Green's functions}

The calculation of $Z_0[j]$ is facilitated by introducing the
Nambu-Gor'kov formalism~\cite{nambu,book:fetter,book:griffin}. Let
us introduce the (bosonic) two-component Nambu-Gor'kov fields
\begin{equation}
\Psi(\xt)=
\begin{bmatrix}
\chi(\xt) \\
\chi^\dagger(\xt)
\end{bmatrix},\quad \Psi^\dagger(\xt)=
\left[\chi^\dagger(\xt),\chi(\xt)\right], \label{spinorPsi}
\end{equation}
the corresponding sources
\begin{equation}
J(\xt)= \begin{bmatrix}
j(\xt) \\
j^\ast(\xt)
\end{bmatrix},\quad
J^\dagger(\xt)= \left[j^\ast(\xt),j(\xt)\right],\label{spinorJ}
\end{equation}
and the following $2\times 2$ matrices
\begin{equation}
\sigma_+ = \begin{bmatrix}
0 & 1 \\
0 & 0
\end{bmatrix},\quad
\sigma_- = \begin{bmatrix}
0 & 0 \\
1 & 0
\end{bmatrix},\quad
\sigma_3 = \begin{bmatrix}
1 & 0 \\
0 & -1
\end{bmatrix}, \label{matrices}
\end{equation}
in terms of which the quadratic part of the Lagrangian given by
(\ref{Lint}) and now defined on the contour $\mathcal{C}$ in
Fig.~\ref{fig:ctpcontour} can be written as
\begin{eqnarray}
\mathcal{L}_0[\Psi^\dagger,\Psi] &=& \frac{1}{2} \Psi^\dagger
\bigg[i \sigma_3\frac{\partial}{\partial t} +
\frac{\nabla^2}{2m}+\mu-2 g |\phi_0|^2 -g\phi^2_0\sigma_+
-g\phi_0^{\ast 2} \sigma_- \bigg]\Psi, \label{NGlag}
\end{eqnarray}
where the time derivative is understood to be taken with respect
to the contour $\mathcal{C}$. The equation of motion for the free
Nambu-Gor'kov field $\Psi$ in the presence of the source $J$ reads
\begin{equation}
\left[i \sigma_3\frac{\partial}{\partial t} +
\frac{\nabla^2}{2m}+\mu-2 g |\phi_0|^2-g\phi^2_0 \sigma_+
-g\phi^{\ast 2}_0\sigma_-\right]\Psi(\xt)=
-J(\xt).\label{eqnofmot}
\end{equation}

The solution of this equation of motion is given by
\begin{equation}
\Psi_{J}(\xt) = -\int_\mathcal{C} d^4x'
G(\bx-\bx',t-t')J(\bx',t'), \label{sol}
\end{equation}
where $G(\bx-\bx',t-t')$ is the Green's function along this
contour and satisfies
\begin{eqnarray}
&&\left[i \sigma_3\frac{\partial}{\partial t} +
\frac{\nabla^2_\mathbf{x}}{2m}+\mu-2 g |\phi_0|^2-g\phi^2_0
\sigma_+ -g\phi^{\ast 2}_0 \sigma_-\right]\nn\\
&&\qquad\qquad\qquad\qquad\times\,G(\bx-\bx',t-t')=
\delta_\mathcal{C}(t-t')\,\delta^{(3)}(\bx-\bx'),\label{green}
\end{eqnarray}
with $\delta_\mathcal{C}(t-t')$ the Dirac delta function along the
contour $\int_\mathcal{C}dt'\,\delta_\mathcal{C}(t-t')=1$. The
Green's function $G(\bx-\bx',t-t')$ has the form
\begin{equation}
G(\bx-\bx',t-t')=
G^>(\bx-\bx',t-t')\Theta_\mathcal{C}(t-t')+
G^<(\bx-\bx',t-t')\Theta_\mathcal{C}(t'-t),
\label{Gcont}
\end{equation}
where $\Theta_\mathcal{C}(t-t')$ is the step function along the
contour and $G^\gtrless(\bx-\bx',t-t')$ obey the homogeneous
equations of motion.

The periodic boundary conditions on the fields in the path
integral, a result of the trace over bosonic fields in
(\ref{genefunc}), lead to the following boundary condition on the
Green's function
\begin{equation}
\lim_{t_0\to -\infty} G(\bx-\bx',t_0-t')=\lim_{t_0\to -\infty}
G(\bx-\bx',t_0-i\beta-t').\label{pbc}
\end{equation}
Since along the contour $t_0\to -\infty$ is the \emph{earliest}
time and $t_0-i\beta$ is therefore the \emph{latest} time,
(\ref{pbc}) entails
\begin{equation}\label{pbccond}
\lim_{t_0\to -\infty}G^<(\bx-\bx',t_0-t')=\lim_{t_0\to -\infty}
G^>(\bx-\bx',t_0-i\beta-t'),\quad \forall\; t',
\end{equation}
which is the Kubo-Martin-Schwinger (KMS) condition for equilibrium
correlation functions~\cite{book:kadanoff}.

The free field generating functional $Z_0[J]$ is now obtained by
writing
\begin{gather}
\Psi(\bx,t) = \widetilde{\Psi}(\xt)+ \Psi_J(\xt),\nn\\
\Psi^\dagger(\bx,t) = \widetilde{\Psi}^\dagger(\xt)+
\Psi^\dagger_J(\xt),
\end{gather}
which leads to the result
\begin{equation}
Z_0[J] = Z_0[0]\exp\left[-\frac{i}{2} \int_\mathcal{C} d^4x
\int_\mathcal{C}d^4x' J^\dagger (x) G(x-x')J(x')\right],
\label{genfun0}
\end{equation}
where and hereafter $x$ denotes the space-time coordinates $(\xt)$
for simplicity of notation. The source independent term $Z_0[0]$
will cancel between the numerator and the denominator in all
expectation values in (\ref{expval}).

Furthermore, we are interested in computing correlation functions
of \emph{finite real times} which are defined for fields in the
forward ($+$) and backward ($-$) time branches but not in the
imaginary branch. For these real-time correlation functions the
contributions to the generating functional from one source in the
imaginary branch and another source in either the forward or
backward branch vanish by the Riemann-Lebesgue
lemma~\cite{chou,tadpole,noneqlect}, since the time arguments are
infinitely far apart along the contour. Therefore the contour
integrals of the source terms and Green's functions in the
generating functional factorize into a term in which the sources
are those either in the forward and backward branches and another
term in which \emph{both} sources are in the imaginary
branch~\cite{tadpole,noneqlect}. The latter term (with both
sources in the imaginary branch) cancel between numerator and
denominator in expectation values and the only remnant of the
imaginary branch is through the periodic boundary conditions along
the full contour in the Green's function.

Thus the generating functional for real-time correlation functions
simplifies to the following expression, defined solely along the
forward and backward time
branches~\cite{chou,mahan,kine,tadpole,noneqlect},
\begin{eqnarray}
Z[J^{\pm},J^{\dagger \pm}]&=&\exp\Big[i\int_{-\infty}^{+\infty}
d^4 x\big(\mathcal{L}_\mathrm{int}[-i\delta/\delta
J^{\ast+},-i\delta/\delta
J^+]\nn\\
&&-\mathcal{L}_\mathrm{int}[i\delta/\delta
J^{\ast-},i\delta/\delta
J^-]\big)\Big]\exp\bigg\{-\frac{i}{2}\int_{-\infty}^{+\infty}
\int_{-\infty}^{+\infty}d^4xd^4x'\nn\\
&&\times\,\Big[J^{\dagger+}(x)G^{++}(x,x')J^+(x')+
J^{\dagger -}(x)G^{--}(x,x')J^-(x')\nonumber \\
&&-\,J^{\dagger
+}(x)G^{+-}(x,x')J^-(x')-J^{\dagger-}(x)G^{-+}(x,x')
J^+(x')\Big]\bigg\},\nn\\
\label{realtimegen}
\end{eqnarray}
with
\begin{eqnarray}
&&G^{++}(x,x')=G^>(\bx-\bx',t-t')\Theta(t-t')+
G^<(\bx-\bx',t-t')\Theta(t'-t),\nn \\
&&G^{--}(x,x')=G^>(\bx-\bx',t-t')\Theta(t'-t)+
G^<(\bx-\bx',t-t')\Theta(t-t'),\nn\\
&&G^{-+}(x,x')=G^>(\bx-\bx',t-t'),\nn\\
&&G^{+-}(x,x')=G^<(\bx-\bx',t-t'),\label{gpm}
\end{eqnarray}
where now $-\infty \leq t, t' \leq +\infty$ and the superscripts
$+$, $-$ correspond to the sources defined on the forward ($+$)
and backward ($-$) time branches, respectively. An important issue
that must be highlighted at this stage, is that derivatives with
respect to sources in the forward ($+$) time branch correspond to
insertion of operators \emph{pre-multiplying} the density matrix
$\rho$ and derivatives with respect to sources in the backward
($-$) branch correspond to the insertion of operators
\emph{post-multiplying} the density matrix. That this is so is a
consequence of the fact that the density matrix evolves in time as
$U(t,t_0)\rho_0 U^{-1}(t,t_0)$ with $U(t,t_0)$ the time evolution
operator.

These four correlation functions are not independent because of
the identity
\begin{equation}
G^{++}(x,x')+G^{--}(x,x')-G^{+-}(x,x')-G^{-+}(x,x') = 0.
\label{identity}
\end{equation}
The diagonal elements in $G^{++}(x,x')$ are the normal Green's
functions, representing the propagation of single noncondensate
particles, whereas the off-diagonal elements are the anomalous
Green's functions, corresponding to the annihilation and creation
of two noncondensate particles, respectively.

The functions $G^\gtrless(x,x')$, which are solutions of the
homogeneous free field equation of motion, are simply related to
the correlation functions of the free Nambu-Gor'kov fields $\Psi$,
$\Psi^\dagger$. Indeed, taking variational derivatives of the free
field generating functional $Z_0[J]$ with respect to $j^{\ast\pm}$
and $j^{\pm}$ that make up the source $J^{\pm}$, one can easily
show that
\begin{gather}
G^>(x,x')= -i\begin{bmatrix} \langle \chi(x)
\chi^\dagger(x')\rangle & \langle \chi(x) \chi(x')\rangle
\\ \langle \chi^\dagger(x) \chi^\dagger(x')\rangle & \langle
\chi^\dagger(x)\chi(x')\rangle
\end{bmatrix},\nn\\
G^<(x,x')= -i\begin{bmatrix} \langle
\chi^\dagger(x')\chi(x)\rangle & \langle
 \chi(x')\chi(x)\rangle \\ \langle
\chi^\dagger(x')\chi^\dagger(x)\rangle & \langle
\chi(x')\chi^\dagger(x)\rangle
\end{bmatrix},\label{gorkovpm}
\end{gather}
or alternatively
\begin{equation}
G^>_{ab}(x,x')= -i\langle \Psi_a(x)\Psi_b^\dagger(x')\rangle,\quad
G^<_{ab}(x,x')= -i \langle \Psi_b^\dagger(x')\Psi_a(x) \rangle,
\end{equation}
where and hereafter $a,b=1,2$ denote the Nambu-Gor'kov indices.
The expectation values in the expressions above are in the
non-interacting thermal density matrix which corresponds to the
quadratic part of the Lagrangian (Hamiltonian), i.e., the density
matrix that describes free Bogoliubov quasiparticles in thermal
equilibrium at inverse temperature $\beta$.

While the matrix elements $G^\gtrless_{ab}(x,x')$ can be obtained
through the usual Bogoliubov transformation to the quasiparticle
basis~\cite{bogoliubov,book:griffin,book:fetter,shi}, we present
in the Appendix an alternative derivation of these correlation
functions directly from the spinor solutions of the homogeneous
equations of motion. We find that the correlation functions in
(\ref{gorkovpm}) in the continuum limit are given by
\begin{equation}
G^\gtrless_{ab}(x,x')= \int \frac{d^3k}{(2\pi)^3}\,e^{i\bk \cdot
(\bx-\bx')}G^\gtrless_{ab}(k,t-t'),\label{glesserx}
\end{equation}
where
\begin{eqnarray}
G^>_{ab}(k,t-t')&=& -i\big[[1+n_B(\omega_k)] \mathcal{G}_{ab}(k)
e^{-i \omega_k(t-t')}\nn\\
&&+\,n_B(\omega_k)
\overline{\mathcal{G}}_{ab}(k)e^{i \omega_k(t-t')}\big],\nn\\
G^<_{ab}(k,t-t')&=&-i\big[ n_B(\omega_k) \mathcal{G}_{ab}(k) e^{-i
\omega_k(t-t')}\nn\\
&&+\,[1+n_B(\omega_k)] \overline{\mathcal{G}}_{ab}(k)e^{i
\omega_k(t-t')}\big]. \label{glesser}
\end{eqnarray}
In the above expressions, $k\equiv|\bk|$,
$n_B(\omega)=1/(e^{\beta\omega}-1)$ is the Bose-Einstein
distribution function,
\begin{eqnarray}
&&\mathcal{G}(k)= u^2_k
\begin{bmatrix}
1 & -r_k \phi_0/\phi^\ast_0 \\
-r_k \phi^\ast_0/\phi_0 & r_k^2
\end{bmatrix},\nn\\
&&\overline{\mathcal{G}}(k) = u^2_k
\begin{bmatrix}
r_k^2 & -r_k \phi_0/\phi^\ast_0  \\
-r_k \phi^\ast_0/\phi_0  & 1
\end{bmatrix},\label{gmatrices}
\end{eqnarray}
where $u_k$, $r_k$ are given by (\ref{coefs}) in the Appendix, and
$\omega_k$ is the energy of the free Bogoliubov quasiparticles
(see the Appendix)
\begin{equation}
\omega_k=\left[\left(
\frac{k^2}{2m}-\mu+2g|\phi_0|^2\right)^2-\left(g|\phi_0|^2
\right)^2\right]^{1/2}.\label{qpspectrum}
\end{equation}
Using the relation $1+n_B(\omega) = e^{\beta\omega}\,
n_B(\omega)$, one can easily verify the KMS
condition~\cite{book:kadanoff}
\begin{equation}
G^>(k,t-i\beta-t') = G^<(k, t-t').
\end{equation}
Hence, the correlation functions for the fields $\chi^\dagger$,
$\chi$ that will enter in the nonequilibrium perturbative
expansion are completely determined by
(\ref{glesserx})-(\ref{qpspectrum}).

\subsection{Feynman rules}

From the generating functional of nonequilibrium Green's functions
(\ref{realtimegen}), it is clear that the effective interaction
Lagrangian relevant for the nonequilibrium calculations is given
by
\begin{equation}
\mathcal{L}_\mathrm{int}^\mathrm{eff}[\chi^{\dagger
\pm},\chi^\pm]=\mathcal{L}_\mathrm{int}[\chi^{\dagger
+},\chi^+]-\mathcal{L}_\mathrm{int}[\chi^{\dagger
-},\chi^-],\label{efflag}
\end{equation}
where the fields $\chi^{\dagger \pm}$ and $\chi^\pm$ are defined
on the forward ($+$) and backwards ($-$) time branches
respectively. Consequently, this generating functional leads to
the following Feynman rules that define the perturbative expansion
for calculations of nonequilibrium expectation values.
\begin{itemize}
\item[(i)]{There are \emph{two} sets of interaction vertices defined by
$\mathcal{L}_\mathrm{int}^\mathrm{eff}[\chi^{\dagger\pm},\chi^\pm]$:
those in which the fields are in the forward ($+$) branch and
those in which the fields are in the backward ($-$) branch. There
is a relative minus sign between these two types of vertices. }
\item[(ii)]{There are \emph{four} sets of Green's functions
given by (\ref{gpm}) in terms of the normal and anomalous
correlation functions in the form displayed in (\ref{gorkovpm}).
These correlation functions are completely determined by
(\ref{glesserx})-(\ref{qpspectrum}).}
\item[(iii)]{The combinatoric factors are the same as in the equilibrium
or imaginary-time (Matsubara) formulation.}
\end{itemize}

\section{Relaxation of condensate perturbations: An initial value problem}\label{sec:eqnofmot}

The equations of motion for the small amplitude condensate
perturbation $\delta(x)$ induced by the external source $\eta(x)$
is obtained by implementing the \emph{tadpole method} presented in
Refs.~\cite{tadpole,noneqlect}. This method begins by writing the
original Bose fields in the forward ($+$) and backward ($-$) time
branches as
\begin{equation}
\psi^{\pm}(x)= \phi_0+\delta(x) + \chi^{\pm}(x),
\end{equation}
where $\phi_0$ is the homogeneous condensate in the absence of
external source, $\delta(x)$ is the perturbation of the condensate
induced by the external source which vanishes in the absence of
external source and $\chi^{\pm}(x)$ are the noncondensate fields
in the forward ($+$) and backward ($-$) time branches. The
external source $\eta(x)$ are $c$-number fields and hence taken
the same value in both forward and backward branches. The strategy
to obtain the equations of motion for small amplitude condensate
perturbation $\delta(x)$, $\delta^\ast(x)$ is to consider the
\emph{linear}, \emph{cubic} and \emph{quartic} terms in $\chi(x)$,
$\chi^\dagger(x)$ in perturbation theory and impose the
\emph{tadpole condition}
\begin{equation}
\langle \chi^{\pm}(x) \rangle_\eta = \langle \chi^{\dagger\pm}(x)
\rangle_\eta =0 \label{tadpoleqn}
\end{equation}
order by order in the perturbative expansion, but, consistent with
linear response, only keep contributions linear in $\delta(x)$,
$\delta^\ast(x)$.

\subsection{Lowest order (classical) equations of motion}\label{subsec:lowesteom}

To illustrate this technique within the simplest setting and
before embarking on the lengthy calculation at one-loop order, let
us focus on obtaining the equations of motion for $\delta(x)$,
$\delta^\ast(x)$ to lowest order in $g$ by fulfilling the tadpole
condition $\langle \chi^+(0)\rangle_\eta =0$ to lowest order in
the interaction. Since the lowest order interaction terms in
$\mathcal{L}_\mathrm{int}[\chi^\dagger,\chi]$ are those
\emph{linear} in $\chi^\dagger$ and $\chi$ [see (\ref{Lint})], one
obtains
\begin{eqnarray}
&&\int d^4x\,\bigg\{\langle \chi^+(0)\chi^{\dagger +}(x)\rangle
\bigg[\left(i\frac{\partial}{\partial t}+\frac{\nabla^2}{2m}+\mu-2
g |\phi_0|^2\right)\delta(x)-g \phi^{
2}_0\delta^\ast(x)\nn\\
&&\quad+\,\phi_0\left(\mu-g|\phi_0|^2\right)+\eta(x)\bigg]+\langle
\chi^+(0)\chi^+(x)\rangle \bigg[\bigg(-i\frac{\partial}{\partial
t}+\frac{\nabla^2}{2m}+\mu\nn\\
&&\quad-\,2 g |\phi_0|^2\bigg)\delta^\ast(x)-g \phi^{\ast
2}_0\delta(x)+\phi^\ast_0\left(\mu-g|\phi_0|^2\right)
+\eta^\ast(x)\bigg]\nn \\
&&\quad -\,\langle \chi^+(0)\chi^{\dagger -}(x)\rangle
\bigg[\bigg(i\frac{\partial}{\partial t}+\frac{\nabla^2}{2m}+\mu-2
g |\phi_0|^2\bigg)\delta(x)-g \phi^{2}_0\delta^\ast(x)\nn\\
&&\quad+\,\phi_0
\left(\mu-g|\phi_0|^2\right)+\eta(x)\bigg]-\langle\chi^+(0)\chi^-(x)\rangle
\bigg[\bigg(-i\frac{\partial}{\partial
t}+\frac{\nabla^2}{2m}+\mu\nn\\
&&\quad-\,2 g |\phi_0|^2\bigg)\delta^\ast(x)-g \phi^{\ast
2}_0\delta(x)+\phi^\ast_0\left(\mu-g|\phi_0|^2\right)
+\eta^\ast(x)\bigg]\bigg\}= 0,\label{0order}
\end{eqnarray}
which lead to the lowest order equations of motion for
$\delta(\xt)$ and $\delta^\ast(\xt)$
\begin{eqnarray}
&&\left(i\frac{\partial}{\partial t}+\frac{\nabla^2}{2m}+\mu-2 g
|\phi_0|^2\right)\delta(\bx,t)-g \phi^{
2}_0\delta^\ast(\bx,t)\nn\\
&&\qquad\qquad\qquad\qquad\qquad\qquad
+\,\phi_0\left(\mu-g|\phi_0|^2\right)+\eta(\bx,t)=0,\nn\\
&&\left(-i\frac{\partial}{\partial t}+\frac{\nabla^2}{2m}+\mu-2 g
|\phi_0|^2\right)\delta^\ast(\bx,t)-g \phi^{\ast
2}_0\delta(\bx,t)\nn\\
&&\qquad\qquad\qquad\qquad\qquad\qquad
+\,\phi^\ast_0\left(\mu-g|\phi_0|^2\right)+\eta^\ast(\bx,t)=0.
\label{zeroord}
\end{eqnarray}
The equations of motion obtained from the tadpole condition
$\langle \chi^-(0)\rangle_\eta=0$ are the same as those given in
the above expressions. This set of equations is recognized as the
Gross-Pitaevskii equations~\cite{GP,book:griffin,dalfovo99} for
the condensate wave function linearized for a small amplitude
perturbation around a homogeneous condensate $\phi_0$.

In order to generalize the features gleaned from the lowest order
equations of motion to higher orders in the loop expansion, we
rewrite the above equations in the following illuminating form
\begin{gather}
\left(i\frac{\partial}{\partial
t}+\frac{\nabla^2}{2m}+\mu\right)\delta(\bx,t)-
\Sigma^{(0)}_{11}\delta(\bx,t)-\Sigma^{(0)}_{12}
\delta^\ast(\bx,t)+\mathcal{T}[\phi_0,\phi^\ast_0]=-\eta(\bx,t),\nn\\
\left(-i\frac{\partial}{\partial
t}+\frac{\nabla^2}{2m}+\mu\right)\delta^\ast(\bx,t)-
\Sigma^{(0)}_{22}\delta^\ast(\bx,t)-\Sigma^{(0)}_{21}\delta(\bx,t)
+\mathcal{T}^\ast[\phi_0,\phi^\ast_0]=-\eta^\ast(\bx,t),\label{eqsigT}
\end{gather}
where the normal and anomalous self-energies are given by
\begin{equation}
\Sigma^{(0)}_{11}=\Sigma^{(0)}_{22} = 2 g |\phi_0|^2,\quad
\Sigma^{(0)}_{12} = \Sigma^{(0)\ast}_{21}=g \phi^{ 2}_0, \label{sigma120}
\end{equation}
respectively, and the \emph{tadpole}
$\mathcal{T}[\phi_0,\phi^\ast_0]$ denotes terms independent of the
condensate perturbation $\delta$ and $\delta^\ast$
\begin{equation}
\mathcal{T}[\phi_0,\phi^\ast_0]=\phi_0\left(\mu-g|\phi_0|^2\right).
\label{tadpole0}
\end{equation}

Before proceeding further, we note that whereas this lowest order
calculation is fairly straightforward, the equations obtained
above feature the following important aspects that will be shared
also by the higher order calculations.

(i) {The last two lines in the tadpole condition (\ref{0order}),
which are obtained by replacing $\chi^{\dagger +}\to \chi^{\dagger
-}$ and $\chi^+ \to \chi^-$ in the correlation functions, lead to
the same equations of motion as those obtained from the first two
lines. This is a generic feature to all orders in the perturbative
expansion~\cite{tadpole} and is a consequence of the fact that the
coefficients of the correlation functions are local functions of
$t$ which must be the same in both
branches~\cite{tadpole,noneqlect}.}

(ii) {If the homogeneous condensate $|\phi_0|^2 \neq \mu/g$ then
the tadpole $\mathcal{T}[\phi_0,\phi^\ast_0]$ in (\ref{zeroord})
acts as a driving force for $\delta$, which will therefore be
nonvanishing. Hence, the requirement that the departure from
equilibrium $\delta$  vanishes in the absence of external source,
i.e., when $\eta \equiv 0$ at all times, implies that the
homogeneous condensate must fulfill the lowest order
\emph{equilibrium condition}
\begin{equation}
\mu-g|\phi_0|^2=0. \label{equilzeroord}
\end{equation}
In other words if and only if the homogeneous condensate fulfills
the equilibrium condition (\ref{equilzeroord}), the perturbation
$\delta$ will vanish when the external source $\eta$ vanishes at
all times.}

(iii) {Setting $\delta,\,\eta=0$ in the interaction Lagrangian
(\ref{Lint}) the tadpole condition $\langle \chi^+(0)\rangle_\eta
=0$ leads to the equation for the homogeneous condensate
\begin{equation}
\mathcal{T}[\phi_0,\phi^\ast_0]=
\phi_0\left(\mu-g|\phi_0|^2\right)=0,\label{homotad}
\end{equation}
which determines the equilibrium value of the homogeneous
condensate.

Now consider adding an external space-time independent source term
$\eta$, which leads to a space-time independent shift of the
condensate $\phi_0 \to \phi_0+\delta$. With the homogeneous
condensate now being $\phi_0+\delta$ the corresponding equation of
motion is therefore given by
\begin{equation}\label{homotadshif}
\mathcal{T}[\phi_0+\delta,\phi^\ast_0+\delta^\ast]=-\eta.
\end{equation}
Expanding to linear order in $\delta$ and $\delta^\ast$, this
equation becomes
\begin{equation}
\mathcal{T}[\phi_0,\phi^\ast_0]+\frac{\partial
\mathcal{T}[\phi_0,\phi^\ast_0]}{\partial
\phi_0}~\delta+\frac{\partial
\mathcal{T}[\phi_0,\phi^\ast_0]}{\partial \phi^\ast_0}~\delta^\ast
= -\eta, \label{tadshifted}
\end{equation}
which, upon comparing to the first equation of motion in
(\ref{eqsigT}) for a space-time constant perturbation, leads to
the identification
\begin{equation}
\frac{\partial \mathcal{T}[\phi_0,\phi^\ast_0]}{\partial \phi_0} =
\mu-\Sigma^{(0)}_{11},\quad \frac{\partial
\mathcal{T}[\phi_0,\phi^\ast_0]}{\partial \phi^\ast_0}=
-\Sigma^{(0)}_{12}.\label{der1}
\end{equation}
Furthermore, if the condition (\ref{homotad}) is fulfilled, then
using the expressions for the self-energies (\ref{sigma120}) we
obtain
\begin{equation}
\mu=\Sigma^{(0)}_{11}-\frac{\phi^\ast_0}{\phi_0}\Sigma^{(0)}_{12}.\label{HP0}
\end{equation}
Assuming that the condensate $\phi_0$ is real, which can always be
made by a gauge transformation (\ref{gauge}), the above expression
is recognized as the Hugenholtz-Pines relation~\cite{hugenholtz}.

In general, for an arbitrary phase of the condensate, the lowest
order expression for the Hugenholtz-Pines theorem (\ref{HP0})
accounts for the fact that $\Sigma^{(0)}_{12}$ must be complex by
gauge invariance. Indeed, $\Sigma^{(0)}_{11}$ must be invariant
under the gauge transformation (\ref{gauge}) since it multiplies
$\delta$, whereas $\Sigma^{(0)}_{12}$ must transform as an object
with charge $+2$ since it multiplies $\delta^\ast$ and  the full
equation of motion for $\delta$ must transform as $\delta$ itself.
Similarly, $\Sigma^{(0)}_{21}$ must transform as an object with
charge $-2$ under the gauge transformation (\ref{gauge}). This is
an important point that will be taken up again in higher order
calculations and will be the basis for the alternative derivation
of the Hugenholtz-Pines theorem to all orders in perturbation
theory (see Sec.~\ref{sec:HP}).}

While these features emerge as rather trivial relationships that
are gleaned directly from the lowest order (classical) equations
of motion, they will manifest themselves to \emph{all orders} in
the perturbation expansion in the tadpole method. The much less
trivial statement that follows to all orders in the perturbative
expansion implemented via the tadpole method and that will be
confirmed by an explicit calculation below, is that this method
leads to a \emph{gapless} approximation in the classification of
Hohenberg and Martin~\cite{hohenberg}.

Furthermore, the lowest order equilibrium condition
\eqref{equilzeroord} relates the equilibrium condensate at this
order to the chemical potential by $\mu=g|\phi_0|^2$, which is the
lowest order Hugenholtz-Pines relation \eqref{HP0} and hence leads
to a gapless spectrum for free Bogoliubov quasiparticle
excitations in equilibrium condensate
\begin{equation}
\omega_k=\left[\left(
\frac{k^2}{2m}+g|\phi_0|^2\right)^2-\left(g|\phi_0|^2
\right)^2\right]^{1/2},\label{spectrum}
\end{equation}
as can be easily verified from \eqref{qpspectrum}.

\begin{figure}[t]
\begin{center}
\includegraphics[width=3.75in,keepaspectratio=true]{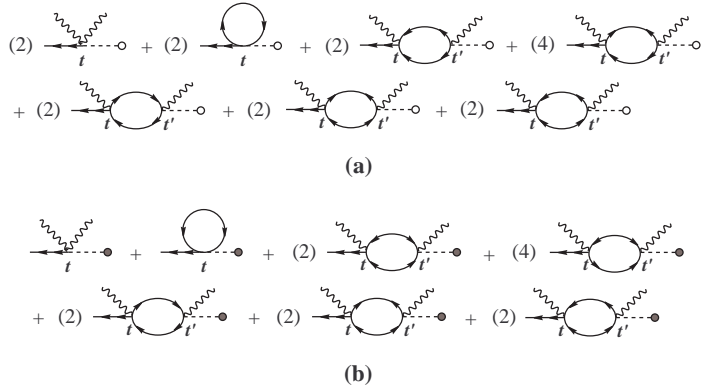}
\caption{Feynman diagrams contributing to the self-energies (a)
$\Sigma^{++}_{11}(x-x')$ and (b) $\Sigma^{++}_{12}(x-x')$ up to
one-loop order with the combinatoric factors denoted in
parenthesis and the labels representing the time branches
suppressed. A solid line with arrows denotes the noncondensate
particle, a wiggly line denotes the condensate $\phi_0$ or
$\phi_0^\ast$, and a dashed line with open (closed) circle denotes
the condensate perturbation $\delta$
($\delta^\ast$).}\label{fig:sigma}
\end{center}
\end{figure}

\subsection{Equations of motion at one-loop order}

The simple example above highlights the tadpole method to obtain
the equations of motion for $\delta(x)$ and $\delta^\ast(x)$,
which we now pursue to one-loop order. The strategy is the same,
but at one-loop order we now face many Feynman diagrams by
treating the effective Lagrangian
$\mathcal{L}_\mathrm{int}^\mathrm{eff}[\chi^{\dagger
\pm},\chi^\pm]$ in perturbation theory with the Feynman rules
described above. Just as in the simpler lowest order example
worked out above, there are \emph{two} types of contributions to
the equations of motion: (i) contributions that are proportional
to $\delta$, $\delta^\ast$, these give the corresponding
self-energies, and (ii) contributions that are \emph{independent}
of $\delta$, $\delta^\ast$. The latter are the higher order
generalization of the tadpole $\mathcal{T}[\phi_0,\phi^\ast_0]$
and its complex conjugate in (\ref{eqsigT}).

\begin{figure}[tb]
\begin{center}
\includegraphics[width=3.75in,keepaspectratio=true]{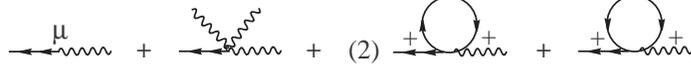}
\caption{Feynman diagrams contributing to the tadpole
$\mathcal{T}[\phi_0,\phi^\ast_0]$ up to one-loop
order.}\label{fig:tadpole}
\end{center}
\end{figure}

Using the Feynman rules described in the previous section, the
tadpole condition $\langle \chi^+(0)\rangle_\eta=0$ leads to the
following expression
\begin{eqnarray}
&&\int d^4x\,\bigg\{\langle \chi^+(0)\chi^{\dagger+}(x)
\rangle\bigg[\left(i\frac{\partial}{\partial
t}+\frac{\nabla^2}{2m}+\mu\right)\delta(x)-\int d^4x'
\big[\Sigma_{11}(x-x')\delta(x')\nn\\
&&\quad+\,\Sigma_{12}(x-x')\delta^\ast(x')\big]+
\mathcal{T}[\phi_0,\phi^\ast_0]+\eta(x)\bigg]+ \langle
\chi^+(0)\chi^{+}(x)\rangle\bigg[\bigg(-i\frac{\partial}{\partial
t}\nn\\
&&\quad+\,\frac{\nabla^2}{2m}+\mu\bigg)\delta^\ast(x)-\int d^4x'
\big[\Sigma_{22}(x-x')\delta^\ast(x')+\Sigma_{21}(x-x')\delta(x')\big]\nn\\
&&\quad+\,\mathcal{T}^\ast[\phi_0,\phi^\ast_0]+\eta^\ast(x)\bigg]-
\left[\chi^{\dagger+}(x)\to\chi^{\dagger-}(x),
\chi^+(x)\to\chi^-(x)\right]\bigg\}=0,\nn\\
\label{eqnofmot1lup}
\end{eqnarray}
where
$\Sigma_{ab}(x-x')=\Sigma^{++}_{ab}(x-x')+\Sigma^{+-}_{ab}(x-x')$
are the self-energies and $\mathcal{T}[\phi_0,\phi^\ast_0]$ is the
tadpole. The diagrams for the self-energies $\Sigma^{++}_{11}$ and
$\Sigma^{++}_{12}$ up to one-loop order are depicted in
Fig.~\ref{fig:sigma} with the labels representing the time
branches suppressed for simplicity, those for $\Sigma^{+-}_{11}$
and $\Sigma^{+-}_{12}$ can be obtained in an analogous manner. The
diagrams for the tadpole $\mathcal{T}[\phi_0,\phi^\ast_0]$ are
depicted in Fig.~\ref{fig:tadpole}.

Introducing the space Fourier transforms for $\delta(x)$,
$\eta(x)$ and $\Sigma_{ab}(x-x')$ as
\begin{equation}
\delta(\xt)=\intk\,\delta_\bk(t)\,e^{i\bk\cdot\bx},~\mbox{etc.},
\end{equation}
we find the final form of the equations of motion in momentum
space to be given by
\begin{eqnarray}
&&\left(i\frac{d}{dt}-\frac{k^2}{2m}+\mu\right) \delta_\bk(t)-
\int_{-\infty}^{+\infty}
dt'\big[\Sigma_{11}(\bk,t-t')\,\delta_\bk(t')+
\Sigma_{12}(\bk,t-t')\nn\\
&&\qquad\times\,\delta^\ast_{-\bk}(t')
\big]+\mathcal{T}[\phi_0,\phi^\ast_0]\,\delta^{(3)}(\bk)+
\eta_\bk(t)=0,\nn\\
&&\left(-i\frac{d}{dt}-\frac{k^2}{2m}+\mu\right)
\delta^\ast_{-\bk}(t)-\int_{-\infty}^{+\infty}
dt'\big[\Sigma_{22}(\bk,t-t')\,\delta^\ast_{-\bk}(t')+
\Sigma_{21}(\bk,t-t')\nn\\
&&\qquad\times\,\delta_\bk(t')\big]+
\mathcal{T}^\ast[\phi_0,\phi^\ast_0]\,
\delta^{(3)}(\bk)+\eta^\ast_{-\bk}(t)=0.\label{eom2}
\end{eqnarray}
The equations of motion obtained from the tadpole condition
$\langle\chi^-(0)\rangle_\eta=0$ are the same as those given in
the above expressions. While the above expression has been
obtained at one-loop order, it is straightforward to conclude
after a simple diagrammatic analysis that the structure of the
equations of motion obtained above is \emph{general and valid to
all orders}.

\begin{figure}[t]
\begin{center}
\includegraphics[width=3.75in,keepaspectratio=true]{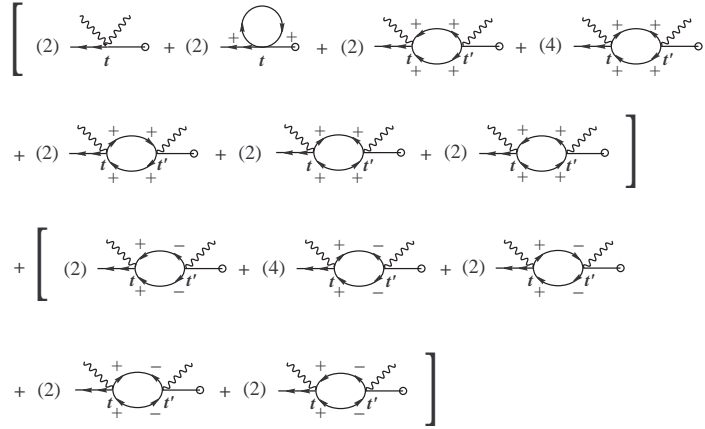}
\caption{Feynman diagrams contributing to the normal self-energy
$\Sigma_{11}(x-x')$. Those for the anomalous self-energy
$\Sigma_{12}(x-x')$ can be obtained from the diagrams in
Fig.~\ref{fig:sigma}(b) by attaching the labels $\pm$ in the same
manner as those depicted in this figure.}\label{fig:sigmatimes}
\end{center}
\end{figure}

Figure~\ref{fig:sigmatimes}, which displays the diagrams for the
normal self-energy $\Sigma_{11}$, clearly shows that the
self-energies have a local, instantaneous contribution (the first
two diagrams in the first bracket), which are known as the
Hartree-Fock-Bogoliubov (HFB) contributions, and a nonlocal,
retarded contribution (the rest of the diagrams) that will lead to
absorptive parts. A simple calculation with the real-time
nonequilibrium Green's functions obtained in the previous section
reveals that
\begin{gather}
\Sigma_{11}(\bk,t-t')=\Sigma^\mathrm{HFB}_{11}\,\delta(t-t')+
\Sigma^R_{11}(k,t-t')\,\Theta(t-t'),\nn\\
\Sigma_{12}(\bk,t-t')=\Sigma^\mathrm{HFB}_{12}\,\delta(t-t')+
\Sigma^R_{12}(k,t-t')\,\Theta(t-t').
\end{gather}
From the explicit expressions for the self-energies or by taking
complex conjugation of the equation of motion for $\delta_\bk(t)$,
we find the properties
\begin{gather}
\Sigma_{21}(\kt-t')=\Sigma^\ast_{12}(-\kt-t'),\nn\\
\Sigma_{22}(\kt-t')=\Sigma^\ast_{11}(-\kt-t'). \label{conjuga}
\end{gather}
Furthermore, rotational and parity invariance imply that the
self-energies are only functions of $k$ (as can be explicitly
confirmed at one-loop order).

The HFB instantaneous parts $\Sigma^\mathrm{HFB}_{11}$ and
$\Sigma^\mathrm{HFB}_{12}$ including the lowest order
contributions (\ref{sigma120}) are given by
\begin{gather}
\Sigma^\mathrm{HFB}_{11}=2g(n_0+\widetilde{n}),\quad
\Sigma^\mathrm{HFB}_{12}=g(\phi^2_0+\widetilde{m}),\nn \\
n_0=|\phi_0|^2,\quad
\widetilde{n}=\langle\chi^\dagger\chi\rangle,\quad
\widetilde{m}=\langle\chi\chi\rangle, \label{sigma1}
\end{gather}
where $n_0$ is the density of the condensate particles,
$\widetilde{n}$ and $\widetilde{m}$ are the one-loop normal and
anomalous densities of the noncondensate particles, respectively.
Using the expressions for the real-time correlation functions
given by (\ref{gorkovpm}) with (\ref{glesser}) and
(\ref{gmatrices}) in the coincidence limit (i.e., $t'\to t$) and
the properties of the Bogoliubov coefficients given by
(\ref{coefs}) in the Appendix, we find
\begin{eqnarray}
\widetilde{n}&=&i\intq\,G^<_{11}(q,0)\nn\\
&=&\intq u_q^2\left[n_B(\omega_q)+
r_q^2[1+n_B(\omega_q)]\right],\nn\\
\widetilde{m}&=&i\intq G^>_{12}(q,0)\nn\\
&=&-\frac{g\phi^2_0}{2}\int\frac{d^3q}{(2\pi)^3\omega_q}
[1+2n_B(\omega_q)],\label{mtilde}
\end{eqnarray}
where in obtaining the final expression of $\widetilde{m}$ use has
been made of (\ref{coefs}) in the Appendix.

As argued above, the normal self-energies $\Sigma_{11}$ and
$\Sigma_{22}$ must be invariant under the gauge transformation
(\ref{gauge}), while the anomalous ones $\Sigma_{12}$ and
$\Sigma_{21}$ must transform as $\phi^2_0$ and $\phi^{\ast 2}_0$,
respectively. Thus using the property (\ref{conjuga}) it is proves
convenient to write
\begin{gather}
\Sigma_{11}(k,t-t')=\Sigma^\ast_{22}(k,t-t')=\Sigma_D(k,t-t'),\nn\\
\Sigma_{12}(k,t-t')=\Sigma^\ast_{21}(k,t-t')=
\left(\phi_0/\phi^\ast_0\right)\Sigma_O(k,t-t'),\label{sigmaDO}
\end{gather}
where both $\Sigma_D$ and$\Sigma_O$ are \emph{invariant} under the
gauge transformation. While rewriting the self-energies in this
manner may seem a redundant exercise, the main point is to
highlight and make explicit their transformation laws under the
gauge transformation. This is an important aspect that needs to be
addressed carefully in order to extract the Ward identities, an
\emph{exact} result of the underlying gauge symmetry to be
explored below.

The gauge invariant HFB instantaneous parts
$\Sigma^\mathrm{HFB}_D$ and $\Sigma^\mathrm{HFB}_O$ can be
obtained straightforwardly from (\ref{sigma1}), whereas the gauge
invariant nonlocal, retarded parts $\Sigma^R_D$ and $\Sigma^R_O$
can be written in terms of their spectral representation as
\begin{eqnarray}
\Sigma_D^R(k,t-t')&=&\intomega\,\big\{
i\,[\overline{S}_{B}(\komega)+\overline{S}_{L}(\komega)]
\cos\omega(t-t')\nn\\
&&+\,[\overline{A}_{B}(\komega)
+\,\overline{A}_{L}(\komega)]\sin\omega(t-t')\big\},\nn\\
\Sigma_O^R(k,t-t')&=&\intomega\,[\widehat{A}_{B}(\komega)+
\widehat{A}_{L}(\komega)]\sin\omega(t-t').\label{sigmasNL}
\end{eqnarray}
The symmetric [$S(\komega)$] and antisymmetric [$A(\komega)$]
spectral functions are even and odd functions of $\omega$,
respectively, and found to be given by (see the Appendix for the
explicit expressions of $u_q$ and $r_q$)
\begin{eqnarray}
\overline{S}_B(\komega)&=&-g^2 n_0\intq\,u_q^2u_p^2\,
[r_p^2(2-4r_q+r_q^2 )-(1-4r_q+2r_q^2)]\nn\\
&&\times\,[1+n_B(\omega_q)+n_B(\omega_p)]
[\delta(\omega-\omega_q-\omega_p)+
\delta(\omega+\omega_q+\omega_p)],\nn\\
\overline{S}_L(\komega)&=&-g^2 n_0\intq\,u_q^2u_p^2\,
[r_p^2(1-4r_q+2r_q^2 )-(2-4r_q+r_q^2)]\nn\\
&&\times\,[n_B(\omega_q)-n_B(\omega_p)]
[\delta(\omega-\omega_q+\omega_p)+
\delta(\omega+\omega_q-\omega_p)],\nn\\
\overline{A}_B(\komega)&=&-g^2 n_0\intq\,u_q^2u_p^2\,
[r_p^2(2-4r_q+r_q^2)+(1-4r_q+2r_q^2)+4r_qr_p]\nn\\
&&\times\,[1+n_B(\omega_q)+n_B(\omega_p)]
[\delta(\omega-\omega_q-\omega_p)-
\delta(\omega+\omega_q+\omega_p)],\nn\\
\overline{A}_L(\komega)&=&-g^2 n_0\intq\,u_q^2u_p^2\,
[r_p^2(1-4r_q+2r_q^2)+(2-4r_q+r_q^2)+4r_pr_q]\nn\\
&&\times\,[n_B(\omega_q)-n_B(\omega_p)]
[\delta(\omega-\omega_q+\omega_p)-
\delta(\omega+\omega_q-\omega_p)],\nn\\
\widehat{A}_B(\komega)&=&-2g^2 n_0\intq\,
u_q^2u_p^2\,[2r_q^2-2r_p(1+r_q^2)+3 r_p r_q]\nn\\
&&\times\,[1+n_B(\omega_q)+n_B(\omega_p)]
[\delta(\omega-\omega_q-\omega_p)-
\delta(\omega+\omega_q+\omega_p)],\nn\\
\widehat{A}_L(\komega)&=&-2g^2 n_0\intq\,u_q^2u_p^2\,
[1+r_p^2r_q^2-2 r_p(1+r_q^2)+3 r_p r_q]\nn\\
&&\times\,[n_B(\omega_q)-n_B(\omega_p)]
[\delta(\omega-\omega_q+\omega_p)-
\delta(\omega+\omega_q-\omega_p)],\label{spectralfunctions}
\end{eqnarray}
where $\bp=\bk+\bq$. The subscript $B$ denotes the Beliaev damping
processes in which one quasiparticle decays into two and the
inverse process~\cite{beliaev,popovld}, and $L$ denotes the Landau
damping processes in which the quasiparticle scatters off another
quasiparticle in the noncondensate~\cite{liu,fed,pita}. The Landau
damping contributions to the nonlocal, retarded parts of the
self-energies arise solely from the thermally excited
quasiparticles and hence vanish in the zero temperature limit. A
careful comparison of the spectral representation of the
self-energies reveals that they coincide with those obtained by
Shi and Griffin~\cite{shi}.

The tadpole term up to one-loop order is given by
\begin{equation}
\mathcal{T}[\phi_0,\phi^\ast_0] =
\phi_0\left[\mu-g\left(n_0+2\widetilde{n}+
\frac{\phi^\ast_0}{\phi_0}\widetilde{m}\right)\right],
\label{tadpole1lup}
\end{equation}
with $\widetilde{n}$ and $\widetilde{m}$ given by (\ref{mtilde}).
Setting the external source $\eta=0$, the equilibrium condition
for the homogeneous condensate $\mathcal{T}[\phi_0,\phi^\ast_0]=0$
leads to
\begin{equation}
\phi_0\left[\mu-g\left(n_0+2\widetilde{n}+\frac{\phi^\ast_0}{\phi_0}
\widetilde{m}\right)\right]=0. \label{equicond}
\end{equation}

In equilibrium and below the critical temperature, the condensate
$\phi_0 \neq 0$, therefore (\ref{equicond}) provides one
relationship between the equilibrium condensate and the chemical
potential. The other relationship is determined by fixing the
total number of particles, i.e.,
\begin{equation}
n=n_0 + \widetilde{n}, \label{numberofparts}
\end{equation}
where $n$ is the total density of the particles. Thus, in
equilibrium the two conditions (\ref{equicond}) and
(\ref{numberofparts}) completely determine the chemical potential
and the equilibrium condensate consistently to one-loop order.

\subsection{Nonequilibrium dynamics as an initial value problem}
In an experimental situation the dynamical evolution of the small
amplitude condensate perturbation is studied by preparing a
Bose-condensed gas slightly perturbed away from equilibrium by
adiabatically coupling to some external source in the infinite
past. Once the source is switched-off at time $t=0$ the perturbed
condensate relaxes towards equilibrium and the relaxation dynamics
is studied. As discussed above, this experimental situation can be
realized within the real-time formulation described here by taking
the spatial Fourier transform of external source to be of the form
\begin{equation}
\eta_\bk(t) = \eta_\bk\,e^{\epsilon t}\,\Theta(-t),\quad \epsilon
\to 0^+.\label{adsource}
\end{equation}
The $\epsilon$-term serves to switch on the source adiabatically
from $t=-\infty$ so as not to disturb the system too far from
equilibrium in the process. If at $t=-\infty$ the system was in an
equilibrium state, then the condition of equilibrium
(\ref{equicond}) ensures that for $t<0$ there is a solution of the
equations of motion (\ref{eom2}) of the form
\begin{equation}
\delta_{\bk}(t) = \delta_{\bk}(0)\,e^{\epsilon t}\quad
\mathrm{for}\;\;t<0,\label{solless0}
\end{equation}
where $\delta_\bk(0)$ is related to $\eta_\bk$ through the
equations of motion for $t<0$. The advantage of the adiabatic
switching-on of the external source is that the time derivative of
the solution (\ref{solless0}) satisfies $\dot{\delta}_{\bk}(t<0)
\to 0$ as $\epsilon \to 0^+$.

Introducing auxiliary quantities $\Pi^R_{ab}(k,t-t')$ defined as
\begin{equation}
\Sigma^R_{ab}(k,t-t')=\frac{\partial}{\partial
t'}\Pi^R_{ab}(k,t-t'),
\end{equation}
then upon using integration by parts, neglecting terms that vanish
in the adiabatic limit $\epsilon \to 0^+$ and assuming that
$\phi_0$ is the equilibrium condensate and hence
$\mathcal{T}[\phi_0,\phi^\ast_0]=0$, the equations of motion
(\ref{eom2}) for $t>0$ are given by
\begin{eqnarray}
&&\left[i\frac{d}{dt}-\frac{k^2}{2m}+\mu-\Sigma^\mathrm{HFB}_{11}-\Pi^R_{11}(k,0)\right]
\delta_\bk(t)-[\Sigma^\mathrm{HFB}_{12}-\Pi^R_{12}(k,0)]\delta^\ast_{-\bk}(t)\nn\\
&&\qquad+\,\int_0^t
dt'\big[\Pi^R_{11}(k,t-t')\dot{\delta}_\bk(t')+\Pi^R_{12}(k,t-t')
\dot{\delta}^\ast_{-\bk}(t')\big]=0,\nn\\
&&\left[-i\frac{d}{dt}-\frac{k^2}{2m}+\mu-\Sigma^\mathrm{HFB}_{22}-\Pi^R_{22}(k,0)\right]
\delta^\ast_{-\bk}(t)-[\Sigma^\mathrm{HFB}_{21}+\Pi^R_{21}(k,0)]\delta_\bk(t)\nn\\
&&\qquad+\,\int_0^t
dt'\big[\Pi^R_{22}(k,t-t')\dot{\delta}^\ast_{-\bk}(t')+\Pi^R_{21}(k,t-t')
\dot{\delta}_\bk(t')\big]=0.\label{eom2l}
\end{eqnarray}
The above coupled equations of motion for the condensate
perturbation are now in the form of an \emph{initial value
problem} with initial conditions specified at $t=0$ and can be
solved by Laplace transform. Introducing a two-component
Nambu-Gor'kov spinor and its Laplace transform
\begin{eqnarray}
&&\Delta_\bk(t)= \begin{bmatrix}
\delta_\bk(t)\\
\delta^\ast_{-\bk}(t)
\end{bmatrix},\quad
\widetilde{\Delta}_\bk(s)=\begin{bmatrix}
\widetilde{\delta}_\bk(s)\\
\widetilde{\delta}^\ast_{-\bk}(s)
\end{bmatrix},
\end{eqnarray}
where $s$ is the complex Laplace variable with $\mathrm{Re}\,s>0$,
one can write the Laplace transformed equations of motion in a
compact matrix form as
\begin{equation}\label{laplaeqn}
\widetilde{G}^{-1}(k,s)\,\widetilde{\Delta}_\bk(s)=\frac{1}{s}
\big[\widetilde{G}^{-1}(k,s)-\widetilde{G}^{-1}(k,0)\big]
\Delta_\bk(0).
\end{equation}
In the above equation, $\widetilde{G}^{-1}(k,s)$ is the inverse
Green's function (matrix) up to one-loop order expressed in terms
of the Laplace variable $s$
\begin{equation}
\widetilde{G}^{-1}(k,s)=\begin{bmatrix}
is-k^2/2m+\mu-\widetilde{\Sigma}_D(k,s) &
-(\phi_0/\phi^\ast_0)\widetilde{\Sigma}_O(k,s)\\
-(\phi^\ast_0/\phi_0)\widetilde{\Sigma}_O(k,s) &
-is-k^2/2m+\mu-\widetilde{\Sigma}_D(k,-s)
\end{bmatrix},\label{eom3}
\end{equation}
with
\begin{equation}
\widetilde{\Sigma}_D(k,s)=
\Sigma^\mathrm{HFB}_D+\widetilde{\Sigma}^R_D(k,s),\quad
\widetilde{\Sigma}_O(k,s)=\Sigma^\mathrm{HFB}_O+\widetilde{\Sigma}^R_O(k,s),
\end{equation}
where $\widetilde{\Sigma}^R_D(k,s)$ and
$\widetilde{\Sigma}^R_O(k,s)$ are the Laplace transforms of
$\Sigma^R_D(k,t-t')$ and $\Sigma^R_O(k,t-t')$, respectively,
\begin{eqnarray}
&&\widetilde{\Sigma}^R_D(k,s)=\int^{+\infty}_{-\infty}
\frac{dk_0}{k_0-is}\,\big[\overline{S}_B(k,k_0)
+\overline{S}_L(k,k_0)+\overline{A}_B(k,k_0)+\overline{A}_L(k,k_0)\big],\nn\\
&&\widetilde{\Sigma}^R_O(k,s)=\int^{+\infty}_{-\infty}
\frac{dk_0}{k_0-is}\big[\widehat{A}_B(k,k_0)+\widehat{A}_L(k,k_0)\big].
\label{sigmatilde}
\end{eqnarray}
In obtaining (\ref{eom3}), we have made use of the
properties
\begin{equation}
\widetilde{\Sigma}_{22}(k,s)=\widetilde{\Sigma}_{11}(k,-s),\quad
\widetilde{\Sigma}_{21}(k,s)=\left(\phi^\ast_0/\phi_0\right)^2
\widetilde{\Sigma}_{12}(k,s),\label{sigmaproperties}
\end{equation}
which are a result of (\ref{conjuga}) and (\ref{sigmasNL}). The
solution of (\ref{eom3}) reads
\begin{equation}
\widetilde{\Delta}_\bk(s)=\frac{1}{s}\big[1-\widetilde{G}(k,s)\,
\widetilde{G}^{-1}(k,0)\big] \Delta_\bk(0), \label{eom4}
\end{equation}
where
\begin{eqnarray}
\widetilde{G}(k,s)&=&\frac{1}{\widetilde{D}(k,s)}\nn\\
&&\times\begin{bmatrix} is+k^2/2m-\mu+\widetilde{\Sigma}_D(k,-s) &
-(\phi_0/\phi^\ast_0)\widetilde{\Sigma}_O(k,s) \\
-(\phi^\ast_0/\phi_0)\widetilde{\Sigma}_O(k,s)&
-is+k^2/2m-\mu+\widetilde{\Sigma}_D(k,s)
\end{bmatrix},\nn\\
\end{eqnarray}
with the denominator $\widetilde{D}(k,s)$ given by
\begin{eqnarray}
\widetilde{D}(k,s)&=&\left[is-k^2/2m+\mu-\widetilde{\Sigma}_D(k,s)\right]
\left[is+k^2/2m-\mu+\widetilde{\Sigma}_D(k,-s)\right]\nn\\
&&+\,\widetilde{\Sigma}^2_O(k,s).\label{D}
\end{eqnarray}

The real-time evolution of the condensate perturbation
$\Delta_\bk(t)$ with an initial value $\Delta_\bk(0)$ is now
obtained from the inverse Laplace transform
\begin{equation}\label{antilap}
\Delta_\bk(t)= \int_\mathcal{B}\frac{ds}{2\pi i}\,e^{st}\,
\widetilde{\Delta}_\bk(s),
\end{equation}
where the Bromwich contour $\mathcal{B}$ runs parallel to the
imaginary axis in the complex $s$ plane to the right of all the
singularities (poles and cuts) of
$\widetilde{\Delta}_\bk(s)$~\cite{ivp}. We note that there is
\emph{no} isolated pole in $\widetilde{\Delta}_\bk(s)$ at $s=0$
since the residue vanishes.

The Green's functions $\widetilde{G}(k,s)$ obtained above (albeit
expressed in terms of the Laplace variable $s$) agree with those
obtained via the Dyson's equations~\cite{beliaev,shi}. However, as
discussed in detail in Refs.~\cite{beliaev,shi} that in
equilibrium the Beliaev self-energies possess infrared divergences
in the limit $k,s\to 0$ and hence care must be taken in analyzing
the analytic properties of $\widetilde{\Delta}_\bk(s)$.

At zero temperature these infrared divergences are logarithmic,
while at finite temperatures, the Bose-Einstein enhancement factor
enhances the infrared divergences to a power law. At zero
temperature the infrared divergences  that appear in the Beliaev
self-energies were first analyzed by Gavoret and
Nozi\`{e}res~\cite{gavoret}, who showed that they cancel out
exactly in the final expressions for physical quantities to all
orders in perturbation theory. Nepomnyashchy and
Nepomnyashchy~\cite{nepomnyashchy} as well as Popov and
Seredniakov~\cite{popov} have analyzed the infrared divergences at
finite temperature and also concluded that a resummation of the
most infrared divergent diagrams leads to a finite result.

Having set up the description of the nonequilibrium dynamics of
perturbations away from the equilibrium condensate as an initial
value problem, we will defer the analysis of the real-time
evolution and the infrared divergences in the self-energies to a
forthcoming article.

\section{Ward identity and Hugenholtz-Pines theorem}\label{sec:HP}
\subsection{General results}
A straightforward diagrammatic analysis with the Feynman rules
described above reveals that the generic structure of the
equations of motion obtained via the tadpole method remains the
same to all orders in perturbation theory. When combined with the
transformation properties of $\phi_0$, $\delta$ and $\chi$ under
the gauge transformation (\ref{gauge}) this general form of the
equations of motion allows to derive to all orders in perturbation
theory the Ward identity for the tadpole
$\mathcal{T}[\phi_0,\phi^\ast_0]$. As described below, this
identity is a consequence of the underlying gauge symmetry and in
equilibrium leads to an alternative derivation of the
Hugenholtz-Pines theorem.

First, consider the case in which $\delta,\,\eta=0$ the tadpole
condition $\langle \chi^+(0) \rangle_\eta =0$ leads to
\begin{equation}
\mathcal{T}[\phi_0,\phi^\ast_0] =0,
\end{equation}
which is the equilibrium condition for the homogeneous condensate.
For a space-time independent shift of the condensate  $\phi_0 \to
\phi_0+\delta$ induced by a space-time independent source $\eta$,
the tadpole condition now leads to
\begin{equation}
\mathcal{T}[\phi_0+\delta,\phi^\ast_0+\delta^\ast] =-\eta,
\end{equation}
which expanding to linear order in $\delta$ and $\delta^\ast$
becomes
\begin{equation}
\mathcal{T}[\phi_0,\phi^\ast_0]+\frac{\partial
\mathcal{T}[\phi_0,\phi^\ast_0]}{\partial
\phi_0}\,\delta+\frac{\partial
\mathcal{T}[\phi_0,\phi^\ast_0]}{\partial
\phi^\ast_0}\,\delta^\ast=-\eta.
\end{equation}

We now compare this equation with the first equation of motion in
(\ref{eom2}), which has the same generic structure as the full
equation of motion obtained to all orders in perturbation theory,
for space-time independent $\delta$ and $\delta^\ast$. We
recognize
\begin{eqnarray}
\frac{\partial \mathcal{T}[\phi_0,\phi^\ast_0]}{\partial \phi_0}
&= &\mu-\Sigma_{11}(k=0,\omega=0), \nn\\
\frac{\partial \mathcal{T}[\phi_0,\phi^\ast_0]}{\partial
\phi^\ast_0} &=& -\Sigma_{12}(k=0,\omega=0), \label{selftads}
\end{eqnarray}
to \emph{all orders in perturbation theory}. This is obviously an
\emph{all-order} generalization of the lowest order result
\eqref{der1}.

The second important ingredient and which stems from the equations
of motion (\ref{eom2}) is that under a gauge transformation
(\ref{gauge}) the tadpole $\mathcal{T}[\phi_0,\phi^\ast_0]$
transforms just as $\delta$, $\phi_0$ and $\eta$, i.e.,
\begin{equation}\label{tadpoletrans}
\mathcal{T}[e^{i\theta}\phi_0,e^{-i\theta}\phi^\ast_0]
=e^{i\theta}\mathcal{T}[\phi_0,\phi^\ast_0].
\end{equation}
Taking the gauge parameter $\theta$ infinitesimal and comparing
the linear terms in $\theta$ in the above expression, we find to
\emph{all orders in perturbation theory} the Ward identity for the
tadpole
\begin{equation}
\mathcal{T}[\phi_0,\phi^\ast_0] =\frac{\partial
\mathcal{T}[\phi_0,\phi^\ast_0]}{\partial \phi_0} ~\phi_0
-\frac{\partial \mathcal{T}[\phi_0,\phi^\ast_0]}{\partial
\phi^\ast_0} \phi^\ast_0.\label{Ward}
\end{equation}
Therefore combining (\ref{selftads}) and the Ward identity
(\ref{Ward}), we obtain the following Ward identity which is an
\emph{exact} relationship between the tadpole and the
self-energies at zero frequency and momentum
\begin{equation}
\phi_0 \left[\mu-\Sigma_{11}(k=0,\omega=0)\right]
+\phi^\ast_0\Sigma_{12}(k=0,\omega=0)=
\mathcal{T}[\phi_0,\phi^\ast_0].\label{exactrel}
\end{equation}

Above the critical temperature \emph{both} the tadpole and the
condensate vanish, thus the above equation becomes a trivial
identity. However, below the critical temperature $\phi_0 \neq 0$
and in equilibrium $\mathcal{T}[\phi_0,\phi^\ast_0]=0$, the above
Ward identity (\ref{exactrel}) leads to the Hugenholtz-Pines
relation
\begin{equation}
\mu =\Sigma_{11}(k=0,\omega=0)  -\frac{\phi^\ast_0}{\phi_0}
\Sigma_{12}(k=0,\omega=0).\label{HPthm}
\end{equation}
It is customary to choose the condensate to be real by redefining
its phase via the gauge transformation (\ref{gauge}), in which
case (\ref{HPthm}) leads to the familiar form of the
Hugenholtz-Pines theorem. However, for a condensate with an
arbitrary phase, the anomalous self-energy $\Sigma_{12}$
\emph{must be} proportional to $\phi^2_0$ since in the equation of
motion it multiplies $\delta^\ast$, which transforms under gauge
transformations just as $\phi^\ast_0$. This fact can be seen
explicitly at both the lowest and one-loop order in the
expressions for the respective anomalous self-energy in
(\ref{sigma120}) and in (\ref{sigmaDO}), (\ref{sigmasNL}) with the
spectral functions given by (\ref{spectralfunctions}). However,
the product $ \Sigma_{12} \delta^\ast$ must transform just as
$\delta$ or $\phi_0$ therefore the phase $\phi^\ast_0/\phi_0$
cancels the phase of $\phi^2_0$ in $\Sigma_{12}$. In terms of the
\emph{gauge invariant} self-energies $\Sigma_D$ and $\Sigma_O$ the
Hugenholtz-Pines theorem (\ref{HPthm}) becomes the more familiar
form
\begin{equation}
\mu
=\Sigma_D(k=0,\omega=0)-\Sigma_O(k=0,\omega=0).\label{famHPthm}
\end{equation}

We emphasize that while the Ward identity (\ref{exactrel}) is an
exact relationship valid in or out of equilibrium (corresponding
to when the tadpole $\mathcal{T}[\phi_0,\phi^\ast_0]$ vanishes or
not, respectively), the Hugenholtz-Pines theorem (\ref{HPthm}) or
alternatively (\ref{famHPthm}) is \emph{only} valid in equilibrium
when the tadpole vanishes. This observation will become important
when we discuss nonequilibrium issues below.

\subsection{Confirmation at one-loop order}
We now show explicitly that the Ward identity (\ref{Ward}) is
fulfilled to one-loop order. The tadpole up to one-loop order is
given by (\ref{tadpole1lup}). The Hartree-Fock-Bogoliubov (local)
contributions to the one-loop self-energies are momentum and
frequency independent and given by (\ref{sigma1}), leading to the
relation
\begin{equation}
\Sigma^\mathrm{HFB}_{11}-\frac{\phi^\ast_0}{\phi_0}
\Sigma^\mathrm{HFB}_{12}=
g\left(n_0+2\widetilde{n}-\frac{\phi^\ast_0}{\phi_0}
\widetilde{m}\right),\label{HFBrel}
\end{equation}
which obviously \emph{do not} fulfill the Ward identity
(\ref{Ward}) with the tadpole up to one loop given by
(\ref{tadpole1lup}). This is a known fact that the
Hartree-Fock-Bogoliubov approximation for the one-loop
self-energies violates the Hugenholtz-Pines theorem.

However, using the explicit form of the nonlocal contributions for
the self-energies given by (\ref{sigmatilde}) with the spectral
functions given by (\ref{spectralfunctions}), it is a tedious but
straightforward exercise to find that (after analytic continuation
to real frequency $\omega$ through the substitution
$s\to-i\omega+0^+$~\cite{tadpole,ivp})
\begin{equation}
\Sigma^R_{11}(k=0,\omega=0)-
\frac{\phi^\ast_0}{\phi_0}\Sigma^R_{12}(k=0,\omega=0)=
2g\frac{\phi^\ast_0}{\phi_0}\widetilde{m},\label{NLrels}
\end{equation}
which is an \emph{exact} relationship up to one-loop order,
obtained for arbitrary $\mu$ and $\phi_0$. This is a purely
algebraic relation which we obtained without the use of either the
dispersion relation of quasiparticles or of the lowest order
relationship between $\phi_0$ and $\mu$ implied by
(\ref{equilzeroord}). Furthermore, we have checked explicitly that
the above relation is independent of the order of limits (i.e.,
$k\to 0$ is taken either before or after $\omega\to 0$). While the
individual contributions feature strong infrared divergences in
the limit $k,\omega \to 0$ at finite temperature (see
Ref.~\cite{shi} for a discussion), they cancel exactly in the
combination (\ref{NLrels}). This cancellation is a manifestation
of the general results of Gavoret and Nozi\`{e}res~\cite{gavoret}
at least for the combination that enters in the Ward identity.

When combined with (\ref{HFBrel}) the above relation
(\ref{NLrels}) leads to
\begin{equation}
\Sigma_{11}(k=0,\omega=0)-\frac{\phi^\ast_0}{\phi_0}
\Sigma_{12}(k=0,\omega=0)=
g\left(n_0+2\widetilde{n}+\frac{\phi^\ast_0}{\phi_0}
\widetilde{m}\right).\label{totrels}
\end{equation}
Collecting the above results, we find up to one-loop order that
\begin{eqnarray}
\frac{\mathcal{T}[\phi_0,\phi^\ast_0]}{\phi_0}&=&
\mu-\left[\Sigma_{11}(k=0,\omega=0)-\frac{\phi^\ast_0}
{\phi_0}\Sigma_{12}(k=0,\omega=0)\right]\nn\\
&=&\mu- g\left(n_0+2\widetilde{n}+\frac{\phi^\ast_0}{\phi_0}
\widetilde{m}\right).\label{WI1lup}
\end{eqnarray}
Therefore, the Ward identity (\ref{Ward}) is manifestly fulfilled
up to one-loop order by including the nonlocal parts of the
self-energies. The equilibrium condition
$\mathcal{T}[\phi_0,\phi^\ast_0]=0$ with the relation
(\ref{totrels}) for the one-loop self-energies shows that the
Hugenholtz-Pines condition (\ref{HPthm}) is fulfilled by properly
including the \emph{nonlocal} parts of the self-energies. This in
turn implies a gapless spectrum for quasiparticle
excitations~\cite{hugenholtz,griffin96,book:griffin}.

This is an important advantage of the tadpole method of
nonequilibrium field theory: the equations of motion obtained at a
given order in the loop expansion are causal and guaranteed to
fulfill the corresponding Ward identities to the given order. In
the classification of Hohenberg and Martin~\cite{hohenberg} the
tadpole method leads, therefore, to a \emph{gapless} approximation
that fulfills the Hugenholtz-Pines theorem.

\section{Issues in and out of Equilibrium}\label{sec:beyond}
\subsection{Issues in equilibrium: nature of the perturbative expansion}

A relevant question that emerges at this stage is the nature of
the perturbative expansion. Taking the perturbative expansion in
terms of the coupling $g$ (or alternatively in terms of the
T-matrix~\cite{stoof:ft,shi}) there seems to be an inconsistency
in the calculation: the HFB local contributions to the
self-energies as well as all of the tadpole contributions are
formally of order $\mathcal{O}(g)$, while the nonlocal
contributions are formally of order $\mathcal{O}(g^2)$. Obviously
there seems to be a mismatch in the orders of coupling constant,
but from the confirmation of the Ward identity at one-loop order,
this mismatch is \emph{required} to fulfill the Ward identity.

The failure of the perturbative expansion in terms of $g$ can also
be seen as follows. Using the lowest order Hugenholtz-Pines
relation $\mu -gn_0=0$ [see (\ref{equilzeroord})] so that for
fixed chemical potential, according to the grand-canonical
formulation, $gn_0 =\mu \sim \mathcal{O}(1)$. The nonlocal
contributions to the self-energies [see (\ref{spectralfunctions})]
are all of order $g(gn_0)$, therefore of order $\mathcal{O}(g)$,
hence of the same order as the HFB contributions and the tadpole.
This mismatch of powers of coupling constant and the subtleties
associated with this had been recognized in
Refs.~\cite{shi,weichman}.

As emphasized above with the explicit calculation, the framework
that leads to a consistent expansion, at least for temperatures
well below the critical, is the \emph{loop expansion}. The loop
expansion is formally an expansion in powers of $\hbar$ and
\emph{not} of $g$. The basis for the loop expansion is that powers
of $\hbar$ describe quantum corrections, which at finite
temperature are mixed with thermal contributions and can be
systematically implemented as follows.

Firstly the Lagrangian should be formally considered to be
independent of $\hbar$, i.e., \emph{classical}. Although the
coupling $g$ depends on $\hbar$ through its relation with the
$s$-wave scattering length, a systematic implementation of the
loop expansion requires that $\mathcal{L} \sim \mathcal{O}(1)$ in
terms of $\hbar$ and hence $g$ must be understood as a parameter
in the Lagrangian of order $\mathcal{O}(1)$. In the path integral
representation, each path is weighted by the exponential of the
action divided by $\hbar$, so in the path integral the Lagrangian
enters in the form $\sim \exp\left(\frac{i}{\hbar} \int
d^4x\mathcal{L}\right)$. The $\hbar$ can be absorbed in a
redefinition of the quantum fields
\begin{equation}
\chi,\chi^\dagger \to \sqrt{\hbar} \chi, \sqrt{\hbar}
\chi^\dagger.\label{hbarredef}
\end{equation}
The path integral measure is multiplied by an overall factor which
cancels out in all correlation functions.

Writing the Lagrangian as in (\ref{L1}) and gathering the terms
linear, quadratic, cubic and quartic in $\chi$, $\chi^\dagger$ in
$\mathcal{L}_\mathrm{int}$ as
\begin{equation}
\mathcal{L}_\mathrm{int}=
\mathcal{L}_\mathrm{int}^{(1)}+\mathcal{L}_\mathrm{int}^{(2)}+
\mathcal{L}_\mathrm{int}^{(3)}+\mathcal{L}_\mathrm{int}^{(4)},
\end{equation}
then upon rescaling the fields as in (\ref{hbarredef}) one finds
\begin{equation}
\frac{\mathcal{L}}{\hbar} \to \mathcal{L}_0+
\frac{1}{\sqrt{\hbar}}\mathcal{L}_\mathrm{int}^{(1)}+
\mathcal{L}_\mathrm{int}^{(2)}+
\sqrt{\hbar}\mathcal{L}_\mathrm{int}^{(3)}+
\hbar\mathcal{L}_\mathrm{int}^{(4)}.\label{rescalelag}
\end{equation}
It is now straightforward to see that (i) the lowest order term in
the equation of motion (\ref{zeroord}) are of order
$\mathcal{O}(1/\sqrt{\hbar})$ since they arise from the linear
term  $\mathcal{L}_\mathrm{int}^{(1)}$ in the Lagrangian, (ii) the
tadpole and the HFB contributions to the one-loop self-energies
are obtained from $\sqrt{\hbar}\mathcal{L}_\mathrm{int}^{(3)}$ and
are therefore of order $\mathcal{O}(\sqrt{\hbar})$, and (iii) the
nonlocal contributions are obtained by one insertion of
$\mathcal{L}_\mathrm{int}^{(3)}$ and one insertion of
$\mathcal{L}_\mathrm{int}^{(2)}$ and are therefore \emph{also} of
order $\mathcal{O}(\sqrt{\hbar})$. Hence, the tadpole, HFB and
nonlocal contributions to the self-energies are \emph{all} of
order $\mathcal{O}(\sqrt{\hbar})$, therefore $\mathcal{O}(\hbar)$
as compared to the lowest order (classical) contributions. As a
result of this counting, diagrams with $n$-loops are multiplied by
$\hbar^n$ with respect to the classical terms in the equations of
motion. Thus when the tadpole and the self-energies are computed
to the same order in the $\hbar$ expansion the Ward identity is
guaranteed to be fulfilled. While the identification of the loop
expansion with a power series expansion in $\hbar$ is well known
in the noncondensed phase, the presence of new interaction
vertices proportional to the condensate $\phi_0$ in the condensed
phase introduces different powers of $\hbar$ in the rescaled
Lagrangian. The necessity for replacing the expansion in terms of
the coupling $g$ by a loop expansion in terms of quantum (and
thermal) corrections was first recognized in Ref.~\cite{weichman},
but to our knowledge the consistency of the loop expansion for
fulfilling the Ward identity has not been highlighted in detail
before.

The consistency of the loop expansion still requires the
resolution of another important issue, which has already been
recognized in Ref.~\cite{weichman}. In equilibrium and for a fixed
chemical potential, the equilibrium value of the condensate is
determined by the equilibrium condition
$\mathcal{T}[\phi_0,\phi^\ast_0]=0$. As argued above, the
expression for the tadpole $\mathcal{T}[\phi_0,\phi^\ast_0]$ can
be recognized as the $\delta$ independent terms in the equations
of motion in the loop expansion. Since the tadpole is obtained as
a power expansion in $\hbar$, the equilibrium value of the
condensate $\phi_0$ can also be found as an expansion in $\hbar$.
However, in the diagonalization of the quadratic part of the
Lagrangian $\mathcal{L}_0$, the condensate $\phi_0$ enters as a
parameter. As a result, the Bogoliubov coefficients $u_k$, $v_k$
and the dispersion relation of quasiparticles $\omega_k$ depend
implicitly on $\hbar$ through $\phi_0$ (see the Appendix). Thus
keeping the full $\phi_0$ in the free quasiparticle Lagrangian
$\mathcal{L}_0$ and eventually identifying $\phi_0$ with the
solution of the equilibrium condition
$\mathcal{T}[\phi_0,\phi^\ast_0]=0$ in the loop expansion, leads
to an inconsistent expansion in $\hbar$.

The remedy for this problem is to expand the full $\phi_0$ in
powers of $\hbar$
\begin{equation}
\phi_0 = \overline{\phi}_0+ \Delta,\label{quancorfi0}
\end{equation}
where $\overline{\phi}_0 \sim \mathcal{O}(\hbar^0)$ and $\Delta=
\hbar \Delta^{(1)}+ \hbar^2 \Delta^{(2)}+\cdots$. Thus in the free
quasiparticle Lagrangian $\mathcal{L}_0$ we now replace $\phi_0$
by the zeroth order term $\overline{\phi}_0$, while the
contributions that include $\Delta$ are now lumped together in the
interaction Lagrangian $\mathcal{L}_\mathrm{int}$. The new
interaction Lagrangian now becomes
\begin{eqnarray}
\mathcal{L}_\mathrm{int}[\chi,\chi^\dagger]&=&\chi^\dagger
\bigg[\left(i\frac{\partial}{\partial t}+\frac{\nabla^2}{2m}+\mu-2
g |\phi_0|^2\right)\delta-g \phi^{
2}_0\delta^\ast+\phi_0\left(\mu-g|\phi_0|^2\right)\nn\\
&&+\,\eta\bigg]-g\left(\overline{\phi}_0\Delta^\ast+\overline{\phi}^\ast_0
\Delta+|\Delta|^2\right)\chi^\dagger \chi
-\frac{g}{2}\left(2\overline{\phi}_0+\Delta^2
\right)\chi^\dagger\chi^\dagger\nn\\
&&-\,2g \phi_0^* \delta\chi^\dagger\chi-g \phi_0\delta
\chi^\dagger\chi^\dagger -g\phi_0\chi
\chi^\dagger\chi^\dagger-g\delta\chi \chi^\dagger\chi^\dagger
-\frac{g}{2} \chi^{\dagger }\chi^{\dagger}\chi\chi\nn\\
&&+\,\mathrm{H.c.},\label{lintnew}
\end{eqnarray}
where $\phi_0=\overline{\phi}_0+\Delta$ in all the terms that
involve $\phi_0$.

In this manner, a perturbative expansion around the equilibrium
condensate now in terms of the free quasiparticle Lagrangian that
only involves $\overline{\phi}_0$ and interactions that are
systematically treated in the loop ($\hbar$) expansion is
consistently cast in terms of the gapless free Bogoliubov
quasiparticles as the noninteracting states. The equation of
motion, now consistently obtained in the loop expansion will
involve a tadpole and self-energies that are expanded to the same
order in $\hbar$, thus guaranteeing that the Ward identity is
fulfilled and in equilibrium the quasiparticle excitations are
gapless by the Hugenholtz-Pines theorem. A straightforward
calculation reveals that to $\mathcal{O}(\hbar)$ our calculation
is consistent by using $\overline{\phi}_0$ in the Bogoliubov
coefficients and dispersion relations in the propagators. The
resulting tadpole equation can be systematically solved up to
$\mathcal{O}(\hbar)$ and the self-energies are unambiguously of
order $\mathcal{O}(\hbar)$ with respect to the classical term. The
Ward identity is fulfilled exactly in the manner displayed in the
previous section.

Near the critical temperature the loop expansion fails because of
the presence of infrared divergences. Either renormalization
group~\cite{weichman,stoof:rg} or other nonperturbative
techniques~\cite{blaizot,arnold} must be invoked for a systematic
study of both the static and the dynamical aspects near the
critical point, which of course fall outside the scope of this
article.

\subsection{Issues out of equilibrium: condensate instabilities}

When the condensate is in equilibrium, i.e.,
$\mathcal{T}[\phi_0,\phi^\ast_0]=0$, the solution of the equation
of motion for small amplitude perturbations around equilibrium
will reveal the frequencies and damping rates of the quasiparticle
excitations of the condensate. However, the derivation of the
equations of motion allows a more general study and in particular
to address the question of dynamics \emph{away} from equilibrium.
If the tadpole $\mathcal{T}[\phi_0,\phi^\ast_0]\neq 0$, the
homogeneous condensate $\phi_0$ does \emph{not} describe a
situation of equilibrium. This can be seen, for example, from the
equations of motion (\ref{eom2}), which even for $\eta_\bk(t)=0$
features a nonvanishing inhomogeneity
$\mathcal{T}[\phi_0,\phi^\ast_0]$.

To understand the nonequilibrium aspects, let us first focus on
the spectrum of quasiparticle excitations at \emph{lowest} order.
The excitation energies of free Bogoliubov quasiparticles
$\omega_k$ given by (\ref{qpspectrum}) become \emph{imaginary} for
long wavelengths $k \thickapprox 0$ when
\begin{equation}\label{instability}
(\mu-g|\phi_0|^2)(\mu-3g|\phi_0|^2)< 0.
\end{equation}
These imaginary frequencies correspond to instabilities of the
homogeneous Bose gas for the values of chemical potential $\mu$
and condensate $\phi_0$ for which the condition
(\ref{instability}) is fulfilled. These modes cannot be treated as
stable quasiparticles and their treatment has to be modified, in
particular no occupation number can be defined for these modes.
The equations of motion in this situation describes a nonvanishing
force term proportional to $\mathcal{T}[\phi_0,\phi^\ast_0]$ that
makes the long wavelength condensate perturbations to evolve in
time. The instabilities for long wavelength excitations reveal the
growth of the condensate perturbations~\cite{stoofinst,bijlsma}
and are qualitatively similar to spinodal decomposition in
self-interacting field theories~\cite{boylee}, which must be
studied self-consistently and nonperturbatively. This is a topic
that we are currently study with the formulation presented in this
article and on which we expect to report soon.

\section{Conclusions}\label{sec:conclusions}

In this article we have focused on establishing a program to
obtain the equations of motion for small amplitude perturbations
of a homogeneous condensate based on linear response in a manner
consistent with the underlying symmetries. We introduced a method
that combines the Schwinger-Keldysh formulation of nonequilibrium
quantum field theory, the Nambu-Gor'kov formalism of the
Bogoliubov quasiparticle excitations in the condensed phase and a
novel technique, the tadpole method, used in relativistic field
theory to obtain the equations of motion directly in real time and
consistently in perturbation theory. This method automatically
leads to causal equations of motion and allow to extract the Ward
identities associated with the underlying gauge symmetry that are
valid in or out of equilibrium. In equilibrium, as can be
determined by a definite condition, these Ward identities lead to
the Hugenholtz-Pines theorem, therefore this method leads to
equations of motion that guarantee gapless quasiparticle
excitations of the condensate.

Furthermore, this method leads to a formulation of the time
evolution of perturbations away from the equilibrium condensate as
an initial value problem, thereby establishing contact with
potential experimental situations in which a particular initial
state is prepared away from equilibrium. It also highlights that
the consistent perturbative expansion is \emph{not} an expansion
in the coupling, but a loop expansion, at least at low
temperature. The loop expansion can be systematically implemented
and is guaranteed to fulfill the Ward identities order by order in
perturbation theory. We have obtained the equations of motion
consistently to one-loop order and showed explicitly that the
inclusion of nonlocal (absorptive) contributions to the
self-energies that are beyond the Hartree-Fock-Bogoliubov
approximation and correspond to the Beliaev and Landau damping
processes are necessary to the fulfillment of the Ward identities
in or out of equilibrium.

While our focus in this article is to present the formulation that
leads to causal equations of motion which fulfill the Ward
identities and to explore some of their consequences, our more
overarching goal is to study the real-time dynamics of
nonequilibrium states directly in real time both in homogeneous
and inhomogeneous (trapped) condensates. The homogeneous case
studied here may not bear much relevance to the experimental
situation, however, the techniques presented in this article can
be extrapolated almost immediately to the inhomogeneous case of
trapped atomic gases. Even in the presence of a trapping one-body
potential, the underlying gauge symmetry is global, i.e., the
gauge transformations that leave the Lagrangian invariant are
space independent. However, an important modification will be that
the condensate becomes localized in the center of the trapping
potential and hence breaks translational invariance. This also
implies that the tadpole
$\mathcal{T}[\phi_0(\bx),\phi^\ast_0(\bx)]$ now acquires implicit
space dependence through the inhomogeneous condensate
$\phi_0(\bx)$, which in turn will require a modification of the
steps that lead to the Ward identities. While all of these issues
require a careful study, we expect that the method presented here
will allow us to extract the relevant Ward identities in the
inhomogeneous case as well.

We are currently studying these aspects with particular attention
to the possible infrared divergences and their potential impact on
dynamics away from equilibrium and instabilities of the
condensates. We expect to report on many of these issues in a
forthcoming article.

\section*{Acknowledgements}
The work of D.B.\ and S.-Y.W.\ was supported in part by the US NSF
under grants PHY-9988720, NSF-INT-9815064 and NSF-INT-9905954.
S.-Y.W.\ would like to thank the Andrew Mellon Foundation for
partial support. D.-S.L.\ was supported by the ROC NSC through
grants NSC-90-2112-M-259-010 and NSC-90-2112-M-259-011. H.-L.Y.\
was partially supported by the ROC NSC under grant
NSC-90-2112-M-001-049. H.-L.Y.\ would like to thank LPTHE,
Universit\'e Pierre et Marie Curie (Paris VI) et Denis Diderot
(Paris VII) and the Department of Physics and Astronomy,
University of Pittsburgh for their hospitality. S.M.A.\ would like
to thank King Fahd University of Petroleum and Minerals for
financial support.

\appendix

\section{Plane wave solutions and the Bogoliubov transformation}

In this Appendix we present an alternative derivation of the
correlation functions for $\chi$, $\chi^\dagger$ directly from the
plane wave solutions of the homogeneous equations of motion (i.e.,
in the absence of source) for the Nambu-Gor'kov field given by
(\ref{eqnofmot})
\begin{equation}
\left[i \sigma_3\frac{\partial}{\partial t} +
\frac{\nabla^2}{2m}+\mu-2 g |\phi_0|^2-g\phi^2_0 \sigma_+
-g\phi^{\ast 2}_0\sigma_-\right]\Psi(\xt)= 0.\label{heqnofmot}
\end{equation}
The plane wave solution can be written in the form
\begin{equation}
\Psi(\bx,t) = S(k)\,e^{-i(E_k t- \bk \cdot \bx)},\quad S(k)=
\begin{bmatrix}
U_k \\V_k
\end{bmatrix}, \label{spinorsol}
\end{equation}
where the two-component spinor obeys
\begin{equation}
\left[\begin{array}{cc}
\xi_k & g\phi^2_0 \\
g\phi^{\ast 2}_0 & \xi_k
\end{array}  \right]
\begin{bmatrix}
U_k \\V_k
\end{bmatrix} = E_k\,\sigma_3 \begin{bmatrix}
U_k \\V_k
\end{bmatrix}, \label{matxeqn}
\end{equation}
where $\xi_k = k^2/2m-\mu+2g |\phi_0|^2$. This is an eigenvalue
equation with a ``weight'' $\sigma_3$. The eigenvalues are given
by $E_k=\pm\omega_k$ with $\omega_k = \sqrt{\xi^2_k-(g
|\phi_0|^2)^2}$. The normalization of the positive and negative
energy spinors is chosen so that
\begin{equation}
S^{(\alpha)\dagger}(k)\sigma_3 S^{(\beta)}(k) =
(\sigma_3)_{\alpha\beta},
\end{equation}
where $\alpha,\beta=1,2$ (not to be confused with the components
of the spinors) correspond to $E_k=\pm\omega_k$, respectively.
That this is the correct choice stems from $\sigma_3$ being the
``weight'' in the eigenvalue equation as well as comparing with
the limit $\phi_0=0$ which is the free field case. Introducing the
Bogoliubov coefficients $u_k$ and $v_k$
\begin{gather}
u_k= \left(\frac{\omega_k+\xi_k}{2\omega_k} \right)^{1/2}, \quad
v_k=-u_k\left(\frac{g\phi^{\ast 2}_0}{\omega_k+\xi_k}\right),\nn\\
u_k v_k= -\frac{g\phi^{\ast 2}_0}{2\omega_k},\quad
\frac{v_k}{u_k}= -r_k \frac{\phi^\ast_0}{\phi_0},\quad r_k =
\frac{g|\phi_0|^2}{\omega_k+\xi_k},\label{coefs}
\end{gather}
we find that the positive and negative energy spinors are given by
\begin{equation}
S^{(1)}(k) = \begin{bmatrix}
u_k \\v_k
\end{bmatrix},\quad
S^{(2)}(k) = \begin{bmatrix}
v^\ast_k \\
u_k
\end{bmatrix}, \label{negenspin}
\end{equation}
respectively. After accounting for the interpretation of negative
energy solutions as antiparticles, one can therefore write the
general plane wave solution of the homogeneous equation of motion
(\ref{heqnofmot}) as
\begin{equation}
\Psi(\bx,t) = \frac{1}{\sqrt{V}} \sum_{\bk} \left\{\alpha_\bk
\begin{bmatrix}
u_k \\v_k
\end{bmatrix}e^{-i(\omega_k t-\bk\cdot\bx)} +
\alpha^\dagger_\bk
\begin{bmatrix}
v^\ast_k \\u_k
\end{bmatrix}
e^{i(\omega_k t-\bk\cdot\bx)}\right\}, \label{spinsolu}
\end{equation}
where $V$ is the quantization volume.

At the level of second quantization, one recognizes that
\eqref{spinsolu} is the Bogoliubov transformation. The operators
$\alpha_\bk$ and $\alpha^\dagger_\bk$, respectively, annihilate
and create a Bogoliubov quasiparticle of momentum $\bk$ (energy
$\omega_k$) and obey the usual canonical commutation relations.
The correlation functions of the Nambu-Gor'kov fields in the
density matrix that describes free Bogoliubov quasiparticles in
thermal equilibrium at inverse temperature $\beta$ are therefore
found to be given by
\begin{eqnarray}
\langle \Psi_a(\bx,t) \Psi^{\dagger}_b(\bx',t') \rangle &=&
\frac{1}{V}\sum_{\bk}\Big[[1+n_B(\omega_k)]
\mathcal{G}_{ab}(k)e^{-i \omega_k(t-t')}\nn\\
&&+\,n_B(\omega_k) \overline{\mathcal{G}}_{ab}(k)e^{i
\omega_k(t-t')}\Big]e^{i\bk
\cdot(\bx-\bx')},\nn\\
\langle \Psi^{\dagger}_b(\bx',t') \Psi_a(\bx,t)  \rangle &=&
\frac{1}{V}\sum_{\bk}\Big[n_B(\omega_k) \mathcal{G}_{ab}(k)e^{-i
\omega_k(t-t')}\nn\\
&&+\,[1+n_B(\omega_k)]\overline{\mathcal{G}}_{ab}(k) e^{i
\omega_k(t-t')}\Big]e^{i\bk \cdot (\bx-\bx')}, \label{psidagpsi}
\end{eqnarray}
where $n_B(\omega_k)$ is the equilibrium distribution for
Bogoliubov quasiparticles of momentum $\bk$
\begin{equation}
n_B(\omega_k)=\langle\alpha^\dagger_\bk
\alpha_\bk\rangle=\frac{1}{e^{\beta\omega_k}-1},
\end{equation}
and $\mathcal{G}(k)$, $\overline{\mathcal{G}}(k)$ are given by
(\ref{gmatrices}).


\begin{thebibliography}{99}
\bibitem{bec1}
M.H. Anderson, J.R. Ensher, M.R. Matthews, C.E. Wieman, and E.A.
Cornell, Science \textbf{269}, 198 (1995); D.S. Jin, J. R. Ensher,
M. R. Matthews, C. E. Wieman, E. A. Cornell, Phys. Rev. Lett.
\textbf{77}, 420 (1996); D.S. Jin, M.R. Matthews, J.R. Ensher,
C.E. Wieman, E.A. Cornell, Phys. Rev. Lett. \textbf{78}, 764
(1997).

\bibitem{bec2}
C.C. Bradley, C.A. Sackett, J.J. Tolett, and R.G. Hulet, Phys.
Rev. Lett. \textbf{75}, 1687 (1995); M.-O. Mewes, M.R. Andrews,
N.J. van Druten, D.M. Kurn, D.S. Durfee, and W. Ketterle, Phys.
Rev. Lett. \textbf{77}, 416 (1996).

\bibitem{bec3} K. B. Davis, M.-O. Mewes, M.R. Andrews, N.J.
van Druten, D.S. Durfee, D.M. Kurn, and W. Ketterle, Phys. Rev.
Lett. \textbf{75}, 3969 (1995); M.-O. Mewes, M.R. Andrews, N.J.
van Druten,  D.M. Kurn, D.S. Durfee, C. G. Townsend and W.
Ketterle, Phys. Rev. Lett. \textbf{77}, 988 (1996).

\bibitem{ketterle:rev}
W. Ketterle, D.S. Durfee, and D.M. Stamper-Kurn,
cond-mat/9904034.

\bibitem{snoke} A. Griffin, D.W. Snoke, and S. Stringari,
\emph{Bose-Einstein Condensation} (Cambridge University Press,
Cambridge, 1995).

\bibitem{book:griffin}
A. Griffin, \emph{Excitations in a Bose-Condensed Liquid}
(Cambridge University Press, New York, 1993);
cond-mat/9911419;
cond-mat/9901172.

\bibitem{dalfovo99}
For a recent review, see F. Dalfovo, S. Giorgini, L.P. Pitaevskii,
and S. Stringari, Rev. Mod. Phys. \textbf{71}, 463 (1999).

\bibitem{shi}
H. Shi and A. Griffin, Phys. Rep. \textbf{304}, 1 (1998).

\bibitem{fetter:art}
A. L. Fetter,
cond-mat/9811366.

\bibitem{castin}
Y. Castin,
cond-mat/0105058.

\bibitem{bogoliubov}
N.N. Bogoliubov, J. Phys. (USSR) \textbf{11}, 23 (1947).

\bibitem{bogoexp}
J.M. Vogels, K. Xu, C. Raman, J.R. Abo-Shaeer, and W. Ketterle,
cond-mat/0109205.

\bibitem{GP}
E.P. Gross, Nuovo Cimento \textbf{20}, 454 (1961); J. Math. Phys.
\textbf{4}, 195 (1963); L.P. Pitaevskii, Sov. Phys. JETP
\textbf{13}, 451 (1961).

\bibitem{griffin96}
A. Griffin, Phys. Rev. B \textbf{53}, 9341 (1996).

\bibitem{giorgini98}
S. Giorgini, Phys. Rev. A \textbf{57}, 2949 (1998); \textbf{61},
063615 (2000).

\bibitem{beliaev}
S.T. Beliaev, Sov. Phys. JETP \textbf{34}, 323 (1958).

\bibitem{popovld}
V.N. Popov, Theo. Math. Phys. \textbf{11}, 478 (1972).

\bibitem{liu}
W.V. Liu and W.C. Shieve,
cond-mat/9702122; W.V. Liu, Phys. Rev. Lett. \textbf{79}, 4056
(1997).

\bibitem{fed}
P.O. Fedichev, G.V. Shlyapnikov and J.T.M. Walrave,
cond-mat/9710128.

\bibitem{pita}
L.P. Pitaevskii and S. Stringari, Phys. Lett. A \textbf{235}, 398
(1995).

\bibitem{hohenberg}
P.C. Hohenberg and P.C. Martin, Ann. Phys. (N.Y.) \textbf{34}, 281
(1965).

\bibitem{hugenholtz}
N. Hugenholtz and D. Pines, Phys. Rev. \textbf{116}, 489 (1959).

\bibitem{peskin}
For a pedagogical discussion of Goldstone theorem, see M.E. Peskin
and D.V. Schroeder, \emph{An Introduction to Quantum Field Theory}
(Addison Wesley, Reading, 1995).

\bibitem{gavoret}
J. Gavoret and P. Nozi\`eres, Ann. Phys. (N.Y.) \textbf{28}, 369
(1964).

\bibitem{nepomnyashchy}
Y.A. Nepomnyashchy and A.A. Nepomnyashchy, Sov. Phys. JETP
\textbf{48}, 493 (1978).

\bibitem{popov}
V.N. Popov and A.V. Seredniakov, Sov. Phys. JETP \textbf{50}, 193
(1979).

\bibitem{book:popov}
V.N. Popov, \emph{Functional Integrals and Collective Excitations}
(Cambridge University Press, 1987).

\bibitem{book:fetter}
A. Fetter and J. Walecka, \emph{Quantum Theory of Many-Particle
Systems} (McGraw-Hill, New York, 1971).

\bibitem{stoofinst}
H.T.C. Stoof, Phys. Rev. Lett. \textbf{66}, 3148 (1991); Phys.
Rev. A \textbf{45}, 8398 (1992).

\bibitem{schwinger}
J. Schwinger, J. Math. Phys. \textbf{2}, 407 (1961).

\bibitem{keldysh}
L.V. Keldysh, Sov. Phys. JETP \textbf{20}, 1018 (1965).

\bibitem{kt}
K.T. Mahanthappa, Phys. Rev. \textbf{126}, 329 (1962); P.M. Bakshi
and K.T. Mahanthappa, J. Math. Phys. \textbf{4}, 1 (1963);
\textbf{4}, 12 (1963).

\bibitem{chou}
K.-C. Chou, Z.-B. Su, B.-L. Hao, and L. Yu, Phys. Rep.
\textbf{118}, 1 (1985); J. Rammer and H. Smith, Rev. Mod. Phys.
\textbf{58}, 323 (1986).

\bibitem{mahan}
G. Mahan, \emph{Many Particle Physics}, second edition (Plenum
Press, New York, 1990).

\bibitem{kine}
E. Lifschitz and L. Pitaevskii, \emph{Physical Kinetics} (Pergamon
Press, Oxford, 1981).

\bibitem{stoof:ft}
H.T.C. Stoof, J. Low Temp. Phys. \textbf{114}, 11 (1999);
cond-mat/9910441.

\bibitem{nambu}
Y. Nambu, Phys. Rev. \textbf{117}, 648 (1960); L.P. Gor'kov, Sov.
Phys. JETP \textbf{7}, 505 (1958).

\bibitem{tadpole}
See, for example, D. Boyanovsky, H.J. de Vega, R. Holman, and
D.-S. Lee, Phys. Rev. D \textbf{52}, 6805 (1995); D. Boyanovsky,
M. D'Attanasio, H.J. de Vega, and R. Holman, \textbf{54}, 1748
(1996); and references therein.

\bibitem{ivp}
D. Boyanovsky, H.J. de Vega, D.-S. Lee, Y.J. Ng, and S.-Y. Wang,
Phys. Rev. D \textbf{59}, 105001 (1999); S.-Y. Wang, D.
Boyanovsky, H.J. de Vega, D.-S. Lee, and Y.J. Ng, \textbf{61},
065004 (2000); S.-Y. Wang, D. Boyanovsky, H.J. de Vega, and D.-S.
Lee, \textbf{62}, 105026 (2000).

\bibitem{mikheev}
N.V. Mikheev and N.V. Chistyakov, JETP Lett. \textbf{73}, 642
(2001).

\bibitem{negele}
J.W. Negele and H. Orland, \emph{Quantum Many-Particle Systems}
(Perseus Books, Reading, M.A., 1998).

\bibitem{noneqlect}
For the description of nonequilibrium methods in quantum field
theory see, for example, D. Boyanovsky, H.J. de Vega, and R.
Holman, in \emph{Proceedings of the Second Paris Cosmology
Colloquium}, p.~127, edited by H.J. de Vega and N. Sanchez, (World
Scientific, Singapore, 1995), hep-th/9412052; and references
therein.

\bibitem{book:kadanoff}
L.P. Kadanoff and G. Baym, \emph{Quantum Statistical Mechanics}
(W.A. Benjamin, New York, 1962).

\bibitem{weichman}
P.B. Weichman, Phys. Rev. B \textbf{38}, 8739 (1988).

\bibitem{stoof:rg}
M. Bijlsma and H.T.C. Stoof, Phys. Rev. A \textbf{54}, 5085
(1996).

\bibitem{blaizot}
M. Holzmann, G. Baym, J.-P. Blaizot,
F. Laloe, Phys. Rev. Lett. \textbf{87}, 120403 (2001)

\bibitem{arnold}
P. Arnold and G.D. Moore, Phys. Rev. Lett. \textbf{87}, 120401
(2001).

\bibitem{bijlsma}
M.J. Bijlsma, E. Zaremba, and H.T.C. Stoof, Phys. Rev. A
\textbf{62}, 063609 (2000).

\bibitem{boylee}
D. Boyanovsky, D.-S. Lee, and A. Singh, Phys. Rev. D \textbf{48},
800 (1993); D. Boyanovsky, Phys. Rev. E \textbf{48}, 767 (1993).

\end{thebibliography}
\end{document}